\documentclass[preprint, 5p,12pt,authoryear,times]{elsarticle}

\usepackage{amssymb}

\usepackage{graphicx}
\usepackage{txfonts}
\usepackage[colorlinks=true,linkcolor=blue,citecolor=blue, urlcolor=blue]{hyperref}
\usepackage[justification=centering]{caption}
\usepackage{physics}
\usepackage[separate-uncertainty=true, binary-units]{siunitx}
\usepackage{amsmath}
\usepackage{amssymb}
\usepackage{empheq}
\usepackage{enumitem}
\usepackage{subcaption}
\usepackage{listings}
\usepackage{xcolor}
\usepackage{algorithmic}
\usepackage[ruled,vlined]{algorithm2e}
\usepackage{subcaption}

\DeclareSIUnit \pc {pc}
\DeclareSIUnit \ly {ly}
\DeclareSIUnit \sm {\text{M}_\odot}
\DeclareSIUnit \mag {\text{mag}}
\DeclareSIUnit \airmass {\text{airmass}}
\DeclareSIUnit \px {\text{px}}
\DeclareSIUnit \yr {yr}
\DeclareSIUnit \solum {L_\odot}
\DeclareSIUnit \h {h}
\DeclareSIUnit \hubble {\text{H}_0}
\DeclareSIUnit \jansky {Jy}
\DeclareSIUnit \Jy {Jy}
\DeclareSIUnit \px {px}

\journal{Astronomy and Computing}

\begin{document}

\begin{frontmatter}

\title{Lightweight HI source finding for next generation radio surveys}

\author[epfl,scitas]{Emma Tolley}
\author[epfl]{Damien Korber}
\author[epfl]{Aymeric Galan}
\author[epfl]{Austin Peel}
\author[issi,sussex]{Mark T. Sargent}
\author[epfl]{Jean-Paul Kneib}
\author[epfl]{Frédéric Courbin}
\author[cea]{Jean-Luc Starck}

\affiliation[epfl]{organization={Laboratoire d’Astrophysique, Ecole Polytechnique Federale de Lausanne (EPFL)},
            addressline={Observatoire de Sauverny}, 
            city={Versoix},
            postcode={1290}, 
            country={Switzerland}}

\affiliation[scitas]{organization={SCITAS, Ecole Polytechnique Federale de Lausanne (EPFL)},
            city={Lausanne},
            postcode={1015}, 
            country={Switzerland}}
\affiliation[issi]{organization={International Space Science Institute}, addressline={Hallerstrasse 6}, 
            city={Bern},
            postcode={3012}, 
            country={Switzerland}}

\affiliation[sussex]{organization={Astronomy Centre, Department of Physics \& Astronomy, University of Sussex}, addressline={Falmer}, 
            city={Brighton},
            postcode={BN1 9QH}, 
            country={United Kingdom}}

\affiliation[cea]{organization={Departement d'Astrophysique, CEA-Saclay},
            city={Gif-sur-Yvette},
            postcode={91191}, 
            country={France}}

\begin{abstract}
Future deep HI surveys will be essential for understanding the nature of galaxies and the content of the Universe. However, the large volume of these data will require distributed and automated processing techniques.
We introduce \texttt{LiSA}, a set of python modules for the denoising, detection and characterization of HI sources in 3D spectral data. \texttt{LiSA} was developed and tested on the Square Kilometer Array Science Data Challenge 2 dataset, and contains modules and pipelines for easy domain decomposition and parallel execution. \texttt{LiSA} contains algorithms for 2D-1D wavelet denoising using the starlet transform and flexible source finding using null-hypothesis testing. These algorithms are lightweight and portable, needing only a few user-defined parameters reflecting the resolution of the data. \texttt{LiSA} also includes two convolutional neural networks developed to analyse data cubes which separate HI sources from artifacts and predict the HI source properties. All of these components are designed to be as modular as possible, allowing users to mix and match different components to create their ideal pipeline. We demonstrate the performance of the different components of \texttt{LiSA} on the SDC2 dataset, which is able to find 95\% of HI sources with SNR $> 3$ and accurately predict their properties.

\end{abstract}


\end{frontmatter}


\section{Introduction}
\label{sec:sample1}

Neutral hydrogen (HI) is the most abundant element in our Universe, and observations of its
21-cm (1420.4 MHz) radiation \citep{van1945radiogolven} 
are essential to understand galactic structure, the distribution of dark matter in galaxies, and galactic environmental interactions (\cite{21cmobs}, \cite{21cmgalaxy}).
The sensitivity of current instruments has limited HI survey depths to up to $z\sim 0.5$ \citep{2013pss6.book..183V}. However, the Looking At the Distant Universe with the MeerKAT Array (LADUMA) survey will be able to detect HI emission out to $z\sim1.4$ \citep{blyth2016laduma}, 
and surveys by the Square Kilometer Array Observatory (SKAO)\footnote{ \href{https://astronomers.skatelescope.org/}{https://astronomers.skatelescope.org/}} will be able to push HI surveys even further \citep{2020RSPTA.37890060S}.
As a groundbreaking observational facility, the SKAO introduces new astrophysical, computational, and data analysis challenges.
In particular, the unprecedented amount of imaging data anticipated from SKA extragalactic HI surveys will require fully automated  detection and characterisation of galaxies with minimal manual intervention.

HI survey data is three dimensional: the HI signal emitted by matter is mapped as a function of its position on the sky Doppler-shifted HI 21 cm emission.
While a variety of different tools and techniques exist for radio source detection in 2D imaging data (AEGEAN, \cite{aegean}; ProFound, \cite{profound}; pyBDSF \citep{2015ascl.soft02007M} and 1D spectroscopy 
\citep[GANDALF;][]{gandalf},
source finding in 3D is more complicated and comparatively limited. One popular tool is the \texttt{SoFiA} application \citep{SoFiA}, which provides a selection of algorithms for data filtering and source finding in 3D data. However, SoFiA requires the user definition of over 100 parameters, making it difficult to converge on the correct setting for optimal performance. In the SKA Science Data Challenge 2, over five teams used SoFiA to analyse the same dataset but achieved drastically different results due to different parameter choices.  

Here we present the LIghtweight Source finding Algorithms \texttt{LiSA}\footnote{\texttt{https://github.com/epfl-radio-astro/LiSA}}, a python HI source finding library for next generation radio surveys.
As a python library, \texttt{LiSA} is highly portable and easy to install compared to C++ applications.
\texttt{LiSA} can analyse input data cubes of any size with pipelines that automatically decompose data into different domains for parallel distributed analysis. For source finding, the library contains python modules for wavelet denoising of 3D spatial and spectral data, and robust automatic source finding using null-hypothesis testing. The source-finding algorithms all have options to automatically choose parameters, minimizing the need for manual fine tuning. Finally, \texttt{LiSA} also contains neural network architectures for  classification and characterization of 3D spectral data.

\texttt{LiSA} was developed for and tested on the second SKAO Science Data Challenge\footnote{\texttt{https://sdc2.astronomers.skatelescope.org/}} (SDC2).
In this paper we document the different algorithms contained in \texttt{LiSA}, describe the pipeline developed for the challenge, and show the performance on the SDC2 dataset.

\section{Data}
The SKAO organizes periodic science data challenges to prepare the radioastronomy community for the scale of SKA data and drive the development of new analysis techniques.
SDC2 involved finding and characterizing HI sources in a simulated spectral data cube representing a 2000 hour SKA MID spectral line observation up to $z=0.5$.  The scale of the data is comparable to other HI surveys by SKA precursors such as the DINGO and WALLABY surveys (\cite{askapsurveys}), but with finer spatial resolution reflecting the improved angular resolution of SKA-1 Mid. 

The spectral data cube covers a 20 square degrees area and bandwidth of 950–1150 MHz bandwidth (redshift range 0.235–0.495), with dimensions $5851\times5851\times6668$ voxels in (RA, Dec, frequency). The data have a spatial resolution of 7 arcsec, resulting in a spatial sampling of 2.8 arcsec, and a frequency sampling of 30 kHz. as a consequence, the vast majority of HI sources are nearly unresolved in the spatial dimension but highly resolved in the frequency space. In addition to noise and PSF effects, the data contain artifacts from imperfect continuum subtraction and simulated Radio Frequency Interference (RFI) flagging.

\begin{figure}[h!]
  \centering
    \includegraphics[width=0.5\textwidth]{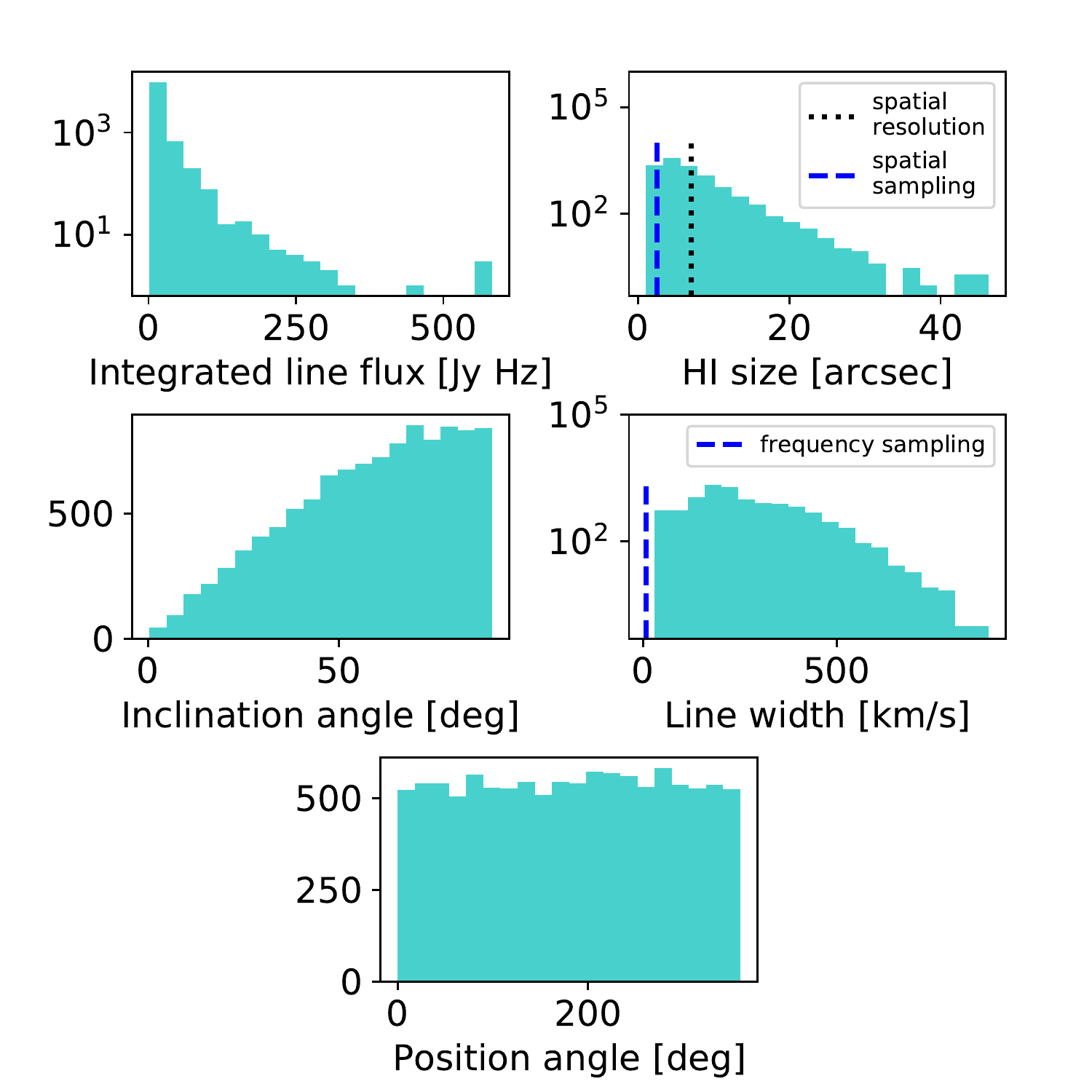}
  \caption{ Properties of HI sources in the development dataset. There are comparatively fewer sources with high line flux integral, HI size, line width, or small inclination angle. \label{fig:var}}
\end{figure}

\begin{figure}[h!]
  \centering
    \includegraphics[width=0.49\textwidth]{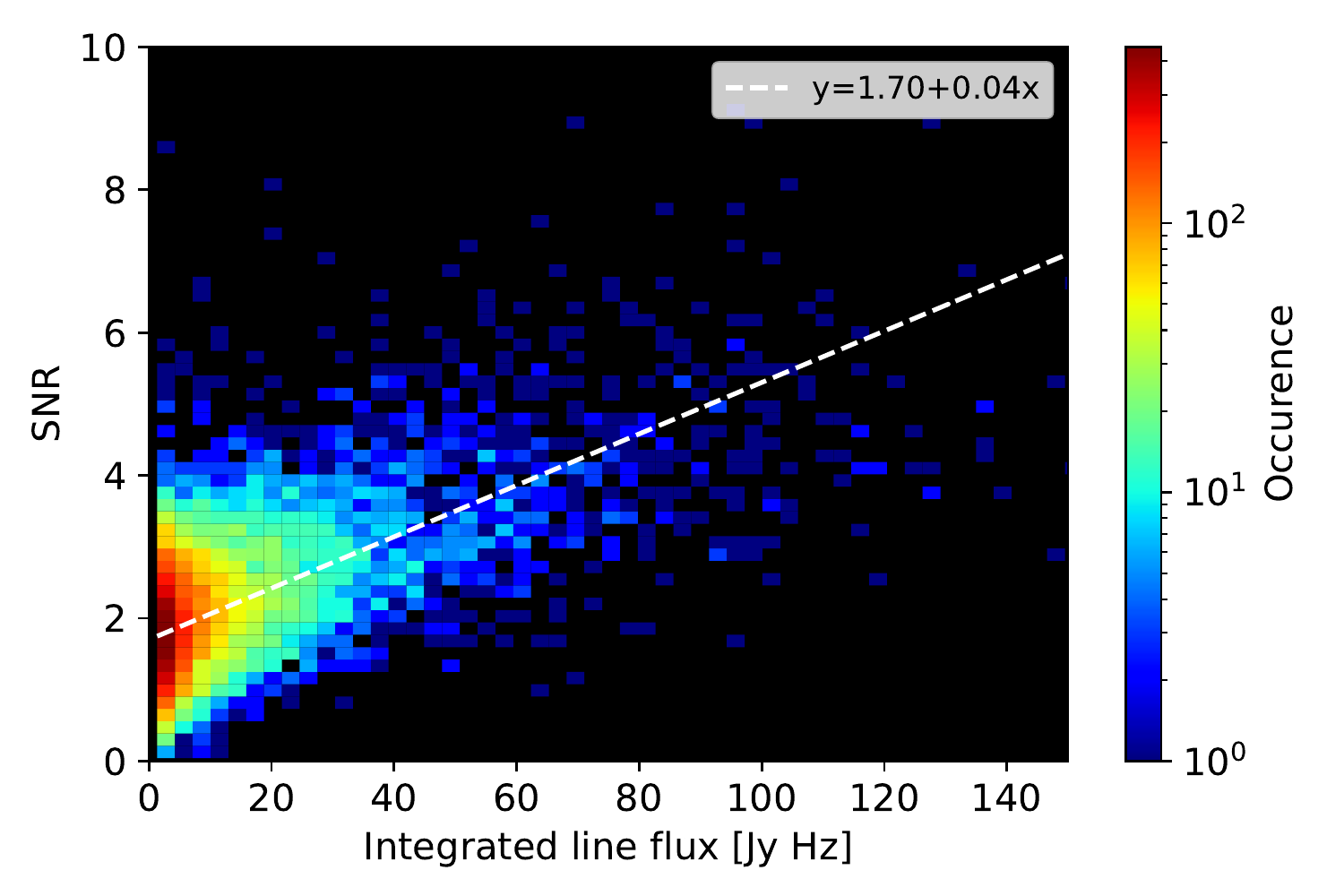}
  \caption{ SNR as a function of integrated line flux for sources in the development dataset. \label{fig:snr}}
\end{figure}

This full dataset was complemented by a smaller development dataset of $1286\times1286\times6668$ voxels and accompanying truth catalog for training and validation purposes. In addition to HI source position, the truth catalog also contained information about the line flux integral, HI size, line width, inclination angle, and position angle of HI source. Distributions for these different variables are shown in Figure~\ref{fig:var}. Most of the sources in the dataset have a very low SNR, as shown in Figure~\ref{fig:snr}.

\section{Domain Decomposition}

\begin{figure*}[t!]
    \centering
    \includegraphics[width=0.85\textwidth]{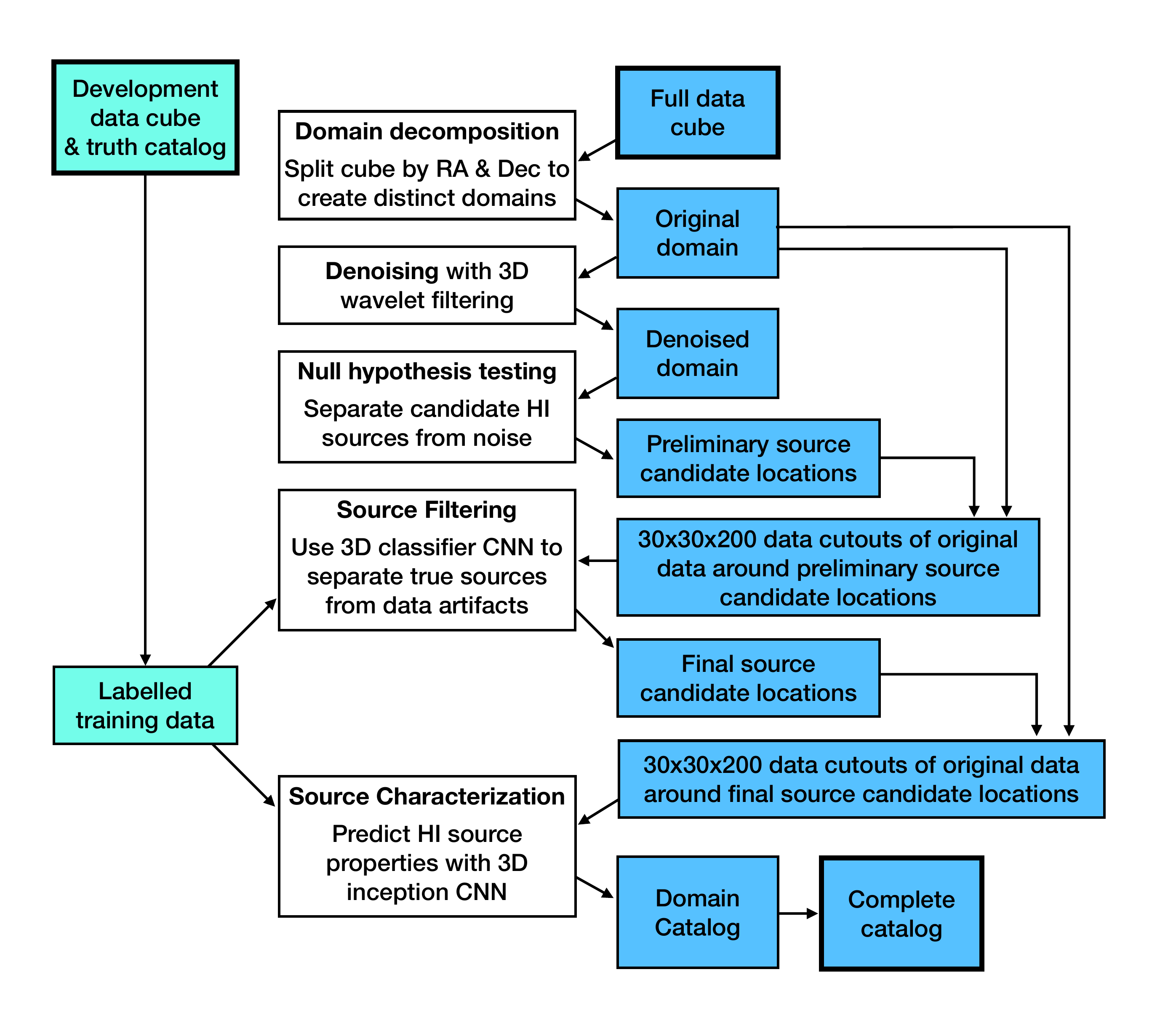}
  \caption{Schematic overview of the \texttt{LiSA} pipeline developed for SDC2. Components of  \texttt{LiSA} are shown in white, and input, output, and intermediate data products are shown in blue. This pipeline 1) splits an input data cube into different domains, 2) denoises the data, 3) uses the denoised data to find source candidates, 4) then uses the source candidate positions to extract subcubes of the original data around each source, 5) filters the sources with a classification CNN, and finally 6) predicts the source properties with a regression CNN. \label{fig:workflow}}
\end{figure*}

At almost 1 TB in size, SDC2 dataset is too large to be read into system memory and must be processed in pieces.
The \texttt{LiSA} pipeline developed for the challenge (shown in Figure~\ref{fig:workflow}) subdivides the input cube into different domains, processes each domain separately, and then concatenates all domains to create the final catalog.

The domain decomposition subdivides the input data along the spatial coordinates and overlaps domains with a configurable border size. The border region is not used for the source finding steps to minimize the double counting of sources at domain edges. However, the border region is used for the source filtering and source characterization steps, so that all of the voxels associated with a source can be used to predict its properties. For running over the SDC2 dataset we used  a border size of 15 voxels, chosen to be twice as large as any of the HI sources in the truth catalog.
To achieve a short turnaround time, we chose to decompose the data cube into 3025 domains, resulting in a domain size of $97\times97\times6668$ voxels.

These domains are provided to other modules in \texttt{LiSA} via the \texttt{DomainReader} interface, through which the algorithms can query or otherwise manipulate the domain data.

\section{Wavelet Denoising}
\subsection{2D-1D wavelet denoising}

We describe a given three-dimensional data cube, or domain, with the following model
\begin{align}
    \label{eq:model_denoising}
    Y = X + N,
\end{align}
where $X$ is the noiseless data cube, and $N$ is a data cube that contains only noise. Our goal is to recover $X$ from the (noisy) observation $Y$, a task known as \textit{denoising}.

Numerous methods for denoising have been developed\footnote{See \citet{Fan2019} for a recent review of denoising methods.}, including techniques based on total variation \citep{Rudin1992}, low-rank minimization \citep[e.g.][]{Ji2010}, and deep learning \citep[e.g.][]{Zhang2017}. One of the most studied and now widely employed methods for denoising a signal is based on wavelets. Wavelet denoising is known to successfully remove noise while preserving features of the true signal regardless of their spatial scales \citep[e.g.][]{Hyeokho1998,Combettes2004}. The method relies on the \textit{wavelet transform} combined with \textit{sparsity} constraints as prior knowledge. Similar to a Fourier transform, a wavelet transform offers a decomposition of a signal with respect to a suitable basis. However, while the Fourier transform only retains frequency information of the original signal, a wavelet transform gives coefficients that encode both the frequency and the spatial location of features. Such a \textit{multi-scale} decomposition, if carefully designed, can lead to a compressed representation of the original signal\footnote{Similar to the Fourier decomposition of a periodic signal.}. A transformed signal that only contains a small number of non-zero coefficients---which is our assumption here---is said to be sparse in wavelet space. Noise, however, is by definition far from being sparse. When a given signal is degraded by noise, many additional non-zero coefficients propagate to the various wavelet scales, leading to a less sparse version of the underlying signal. It is therefore very natural to use sparsity and multi-scale properties of wavelets for denoising, by enforcing the model to be as sparse as possible in wavelet space in order to remove the noise contribution from the observation and to retrieve the underlying signal.

\begin{figure*}[h!]
    \centering
    \includegraphics[width=1\linewidth]{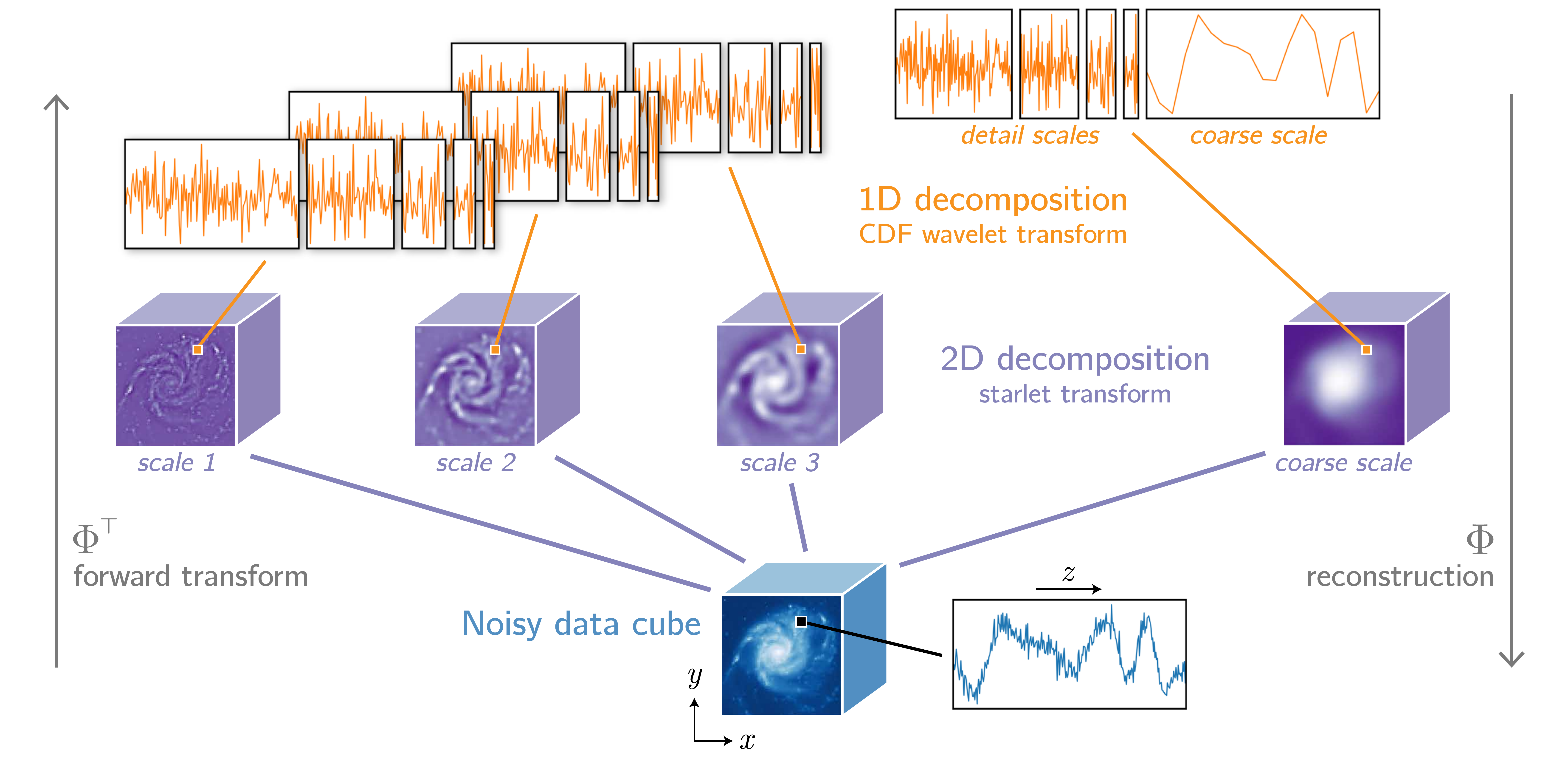}
  \caption{Visualization of the 2D-1D wavelet transform used as a regularization our denoising algorithm. The full transform leads to a multi-scale decomposition in both spatial and spectral dimensions, which is then used in conjunction with a sparsity constraint to allow the efficient separation of true features from noise. \label{fig:wavelet_transform}}
\end{figure*}

We use here the full ``2D + 1D'' information to ensure reliable denoising and to improve the subsequent source detection over the entire domain. Compared to denoising each 2D spatial slice independently, treating the full data cube at once allows us to leverage correlations between spatial and spectral dimensions, which carry important information about the overall shape of the sources. An additional benefit is that the spectral dimension is also denoised in the process. The 2D-1D wavelet transform introduced by \citet{Starck2009} is the optimal choice in our case, which has proven effective at denoising spatio-temporal data \citep{Jiang2015} as well as spatio-spectral data \citep{Floer2012}. We show in Figure~\ref{fig:wavelet_transform} a schematic representation of the 2D-1D transform that illustrates the multi-scale decomposition.

Given that sources exhibit very different spatial morphologies from their spectral ones, different wavelet functions are used in the decomposition of each domain. For the 2D spatial dimension, we use the Isotropic Undecimated Wavelet Transform, also known as the starlet transform \citep{Starck2006,Starck2011}. The starlet transform has been designed to provide sparse representations of astronomical objects that are known to be isotropic or at least composed of many isotropic features that span various spatial scales \citep[e.g. dark matter distributions in galaxy clusters, complex galaxy light distributions,][]{Peel2017,Galan2021}. In the case of unresolved sources, we also expect a sparse representation in starlet space, as the corresponding signal is sparse even in direct space, and is in general close to being isotropic (though to what extent depends on the shape of the PSF). For the 1D spectral dimension, we use the decimated Cohen-Daubechies-Fauveau (CDF) 9/7 wavelet transform \citep{Cohen1992}, which is particularly suited for denoising applications \citep{Vonesch2005}, including in complement to the starlet transform in 2D-1D schemes \citep{Jiang2015}.

\subsection{The inverse problem}

The problem defined by Eq.~\ref{eq:model_denoising} is an inverse problem, where we seek to retrieve $X$ from a noisy observation $Y$. In general, it can be formulated as minimization problem, where the optimal solution $X$ corresponds to the minimum of a loss function. In the case of wavelet denoising, the loss function to be minimized is \citep[e.g.][]{Starck2011}
\begin{align}
    \label{eq:loss_function_denoising}
    L(X) = \frac12 || Y - X ||_2^2 + \lambda ||\Phi^\top X||_1 + \imath_{\geq0}(X),
\end{align}
where the first term on the right-hand side enforces data fidelity, while the remaining two terms serve to regularize the solution by encoding prior knowledge of the solution and avoiding over-fitting. The first regularization term is the sparsity constraint in wavelet space that facilitates the denoising. We also enforce a positivity constraint with the second regularization term. $\Phi^\top$ is the full 2D-1D wavelet transform operator described above, $|X|_1$ is the $\ell_1$-norm that promotes the sparsity of wavelet coefficients, and $\lambda$ is the regularization strength. The positivity constraint is the indicator function $\imath_{\geq0}(X)$ that returns zero if its argument has no negative coefficients and is infinitely large otherwise.

Neither regularization term in Eq.~\ref{eq:loss_function_denoising} is fully differentiable with respect to $X$, preventing us from using simple gradient descent strategies to minimize $L(X)$ directly. Instead, we rely on the formalism of proximal operators \citep{Moreau1962,Combettes2009}, which allows us to combine gradients of the differentiable data-fidelity term with separate corrective operators that apply the regularizations at each step of the minimization. The proximal operator for $\imath_{\geq0}$ is simply a projection onto the positive axis, ${\rm Proj}_{\geq0}(X)$, along each dimension. In contrast, the proximal operator of $\lambda ||\Phi^\top X||_1$ does not have an explicit form, and more complex strategies \citep[e.g. forward-backward algorithms,][]{CombettesWajs2005} are often used in order to retain specific mathematical properties. In our case, we circumvent this issue by considering the proximal operator of $\lambda |X|_1$ instead, which is simply the soft-thresholding operator, ${\rm ST}_\lambda(X)$ \citep{Combettes2009}. This operator shrinks the amplitude of the coefficients by an amount $\lambda$ and sets to zero all coefficients that have an absolute value smaller than $\lambda$. We then approximate the proximal operator of $\lambda ||\Phi^\top X||_1$ by applying the proximal operator of the $\ell_1$-norm to the wavelet coefficients of $X$, namely ${\rm ST}_{\lambda}(\Phi^\top X)$, and apply the reconstruction operator $\Phi$ to recover the domain in direct space. Although there exist iterative algorithms that do not rely on this approximation \citep[e.g.][]{CongVu2011}, we found this approximation to be sufficient for our denoising application.

\subsection{Strength of denoising \label{ssub:denoising_strength}}

The parameter $\lambda$ in Eq.~\ref{eq:loss_function_denoising} effectively controls the strength of the regularization (the thresholding level) and must be well chosen to ensure reliable denoising: a value too high leads to strong denoising with the risk of removing true sources from the data, whereas a value too low leads to incomplete removal of the noise. To objectively set the value of $\lambda$, we use the statistical properties of the data noise that we estimate in wavelet space \citep[e.g.][]{Starck2011}.

Due to the sparse sampling of the visibility space and the process used to convert the observed visibilities to direct space (CLEAN algorithm), the data is expected to be spatially correlated in both spatial and frequency dimensions. We estimate the noise standard deviation in the first 2D scale of the wavelet decomposition, which isolates spatial scales over which the noise dominates the signal. As the noise is varying along the frequency axis, we perform the measurement for each 1D wavelet scale $j$ (i.e. along the frequency axis) independently. We use the median absolute deviation\footnote{The MAD of a signal $x$ is ${\rm MAD}(x) = {\rm median}(\,|\,x - {\rm median}(x)\,|\,)$.} (MAD), which is more robust to outliers compared to the usual standard deviation, and convert it to obtain the Gaussian standard deviation of the noise for the first 2D scale $i=1$, $\sigma_{\Phi}^{(1,j)}=1.48\times{\rm MAD}_{\Phi}^{(1,j)}$. Under the assumption of normally distributed noise, $\sigma_{\Phi}^{(1,j)}$ can be analytically rescaled for each of the remaining 2D wavelet scales $i>1$ to obtain the full estimation of the noise standard deviation $\sigma_{\Phi}^{(i,j)}$ within each 2D and 1D wavelet scale $(i,j)$.

The thresholding level $\lambda$ is then computed independently for each 2D-1D wavelet scale as $\lambda^{(i,j)}=k\,\sigma_\Phi^{(i,j)}$, where $k$ is a scalar that controls the statistical significance of the denoised sources, usually set between 3 and 5 to ensure high-enough detection levels \citep{Starck2011}. In this work, we set $k=5$, leading to denoised domains containing sources detected at the ``5-$\sigma$'' significance level.

We note that the noise estimation algorithm presented above only takes into account correlated noise along the frequency axis. We implemented in \texttt{LiSA} a modified version that takes into account spatially correlated noise, by estimating the noise in each 2D wavelet scale using an iterative scheme. However, we noticed slightly worse results in terms of false source detection rate, hence we do not discuss those results in the present paper.

\subsection{Algorithm}

The steps of our 2D-1D wavelet denoising are summarized in Algorithm~\ref{algo:wavelet_denoising}, that we use to approach the minimum of the loss function in Eq.~\ref{eq:loss_function_denoising}.

\begin{algorithm*}[!h]
\SetAlgoLined
\vspace{0.5em}
\KwData{3D domain extracted from the full data cube ($Y$ in Eq.~\ref{eq:model_denoising})}
\KwResult{Denoised domain}
\SetKwFunction{Denoise}{Denoiser2D1D}
\SetKwProg{Fn}{Function}{:}{}

\vspace{0.5em}
\Fn{\Denoise{Y}}{
Initialize $X$ ; \\
Compute threshold $\lambda$ based on estimated noise levels $\sigma_{\rm \Phi}^{(i,j)}$ (see Sect.~\ref{ssub:denoising_strength}) ; \\
Gradient update (data-fidelity term in Eq.~\ref{eq:loss_function_denoising}): $X \gets X + (Y - X)$ ; \\
Sparsity constraint in wavelet space (second in term in Eq.~\ref{eq:loss_function_denoising}): $X \gets \Phi\left[{\rm ST}_{\lambda}(\Phi^\top X)\right]$ ; \\
Positivity constraint (last in term in Eq.~\ref{eq:loss_function_denoising}): $X \gets {\rm Proj}_{\geq0}(X)$; \\ 
{\bf return} Denoised domain $X$; \\
}
 \caption{2D-1D wavelet denoising algorithm used in \texttt{LiSA}. \label{algo:wavelet_denoising}}
 \vspace{0.5em}
\end{algorithm*}

\vspace{0.5em}
 
Although an iterative version of the algorithm exists, which can, for example, progressively decrease the regularization strength or apply a re-weighting $\ell_1$ scheme to reinforce the sparsity constraint \citep{Candes2007}, we relied on the non-iterative version above, as we found that the denoised cubes were already suitable for the next steps of the pipeline.

This algorithm is similar to that of \citet{Floer2012}, where the authors also implemented a 2D-1D wavelet denoising algorithm and applied it to spectroscopic imaging surveys. There are a few differences with our method, however: (1) they used the starlet transform along the spectral dimension as well as in the spatial domain, (2) they only considered smallest scale of the full wavelet decomposition for detecting compact sources, and (3) they applied the positivity constraint in wavelet space as opposed to direct space. Although it would be interesting to test the effect of their assumptions on our problem, implementing and thoroughly testing different versions of the denoising algorithm is outside the scope of this paper.

\subsection{Denoised domains}

\begin{figure*}[h!]
    \centering
    \includegraphics[width=0.99\textwidth]{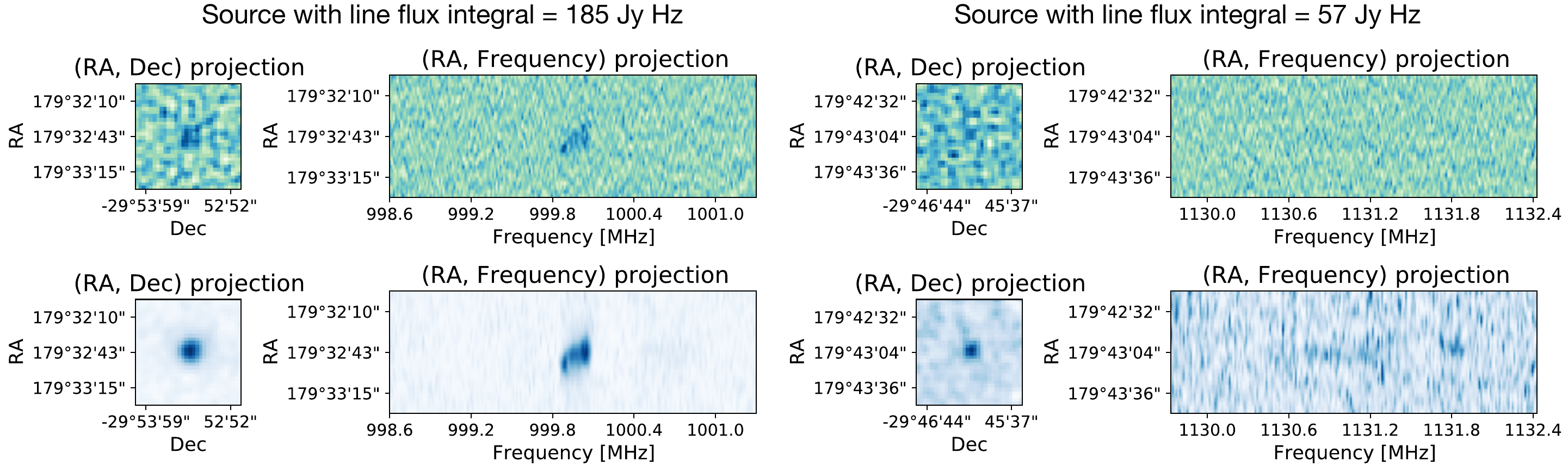}
  \caption{Wavelet denoising applied to two sources, a bright source with an integrated line flux of 185 Jy Hz (left), and a dimmer source with an integrated line flux 57 Jy Hz (right). Projections of the original data are shown in the top row, and projections of the denoised data are shown in the bottom row. \label{fig:denoiseddata}}
\end{figure*}

We show in Figure~\ref{fig:denoiseddata} the result of our denoising method on sources with significantly different line flux integral value (100 and 60 Jy Hz). The denoising procedure significantly improves the signal-to-noise ratio by a factor of 5 for HI sources. However, the denoising causes some amount smoothing of small-scale features, which is expected. Therefore, we do not rely on the denoised data for source characterization steps. The wavelet denoised data is therefore used as an input only to the source-finding step, but not for the source classification or characterization.

\section{Likelihood-based Source Finding}

The goal of the source-finding module is to reliably identify which regions of the data cube correspond to an HI source. Traditional 3D HI source finding algorithms search for local maxima in the dataset which have been smoothed by 3D kernels \citep{2012MNRAS.422.1835S}. However, these strategies are
sensitive to frequency-dependent variations in the noise level.

We developed a null hypothesis testing (NHT) algorithm to find sources in data cubes. This algorithm works by which building a model for the noise in each frequency slice and comparing the model to observed data.

\subsection{Defining the Null Hypothesis}

\begin{figure*}[h!]
  \begin{subfigure}[t]{.5\textwidth}
    \centering
    \includegraphics[width=\textwidth]{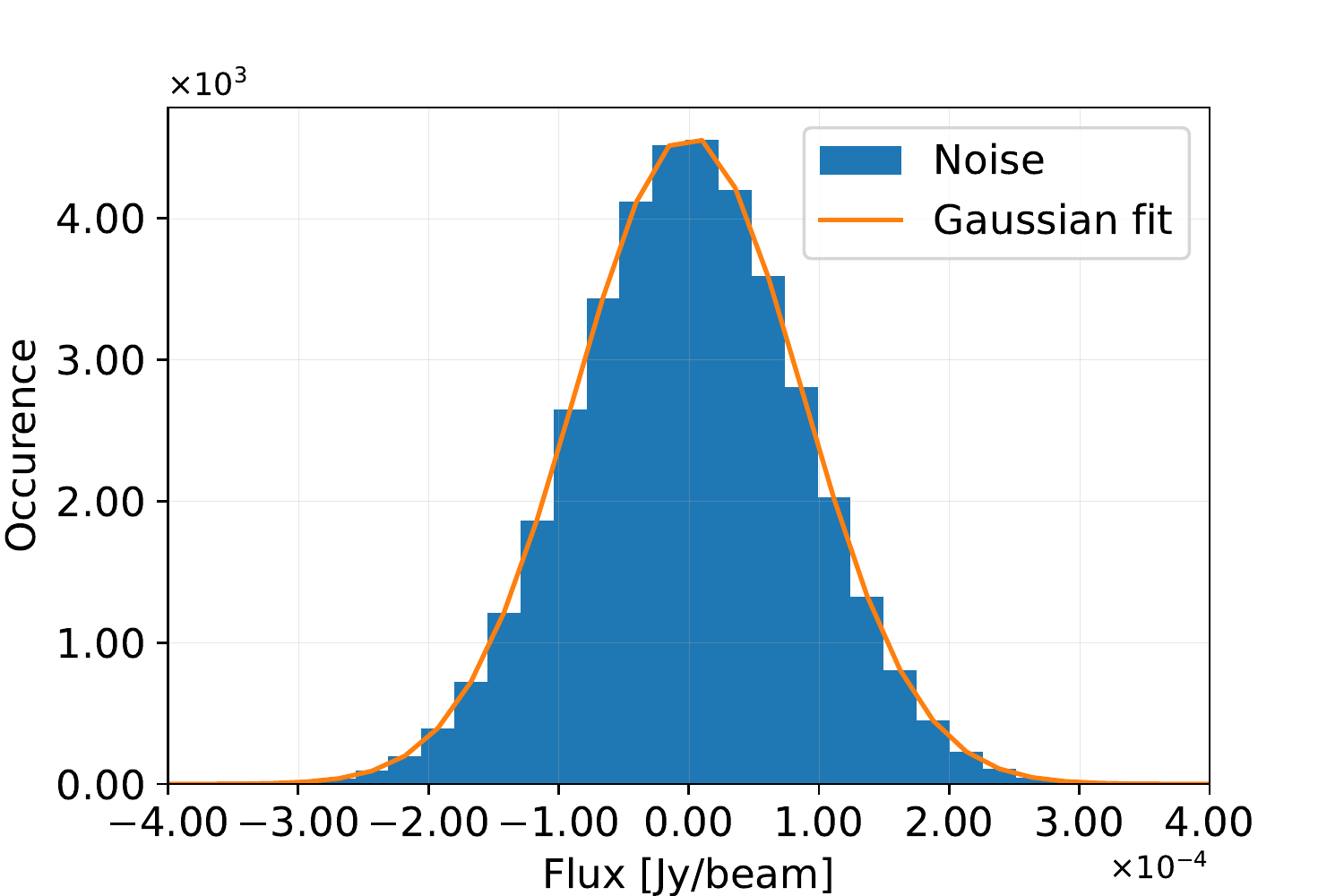}
    \caption{Distribution of noise in one frequency slice of the original (noisy) data cube fitted with Gaussian model.}
    \label{fig:gauss_noise}
  \end{subfigure}
  \begin{subfigure}[t]{.5\textwidth}
    \centering
    \includegraphics[width=\textwidth]{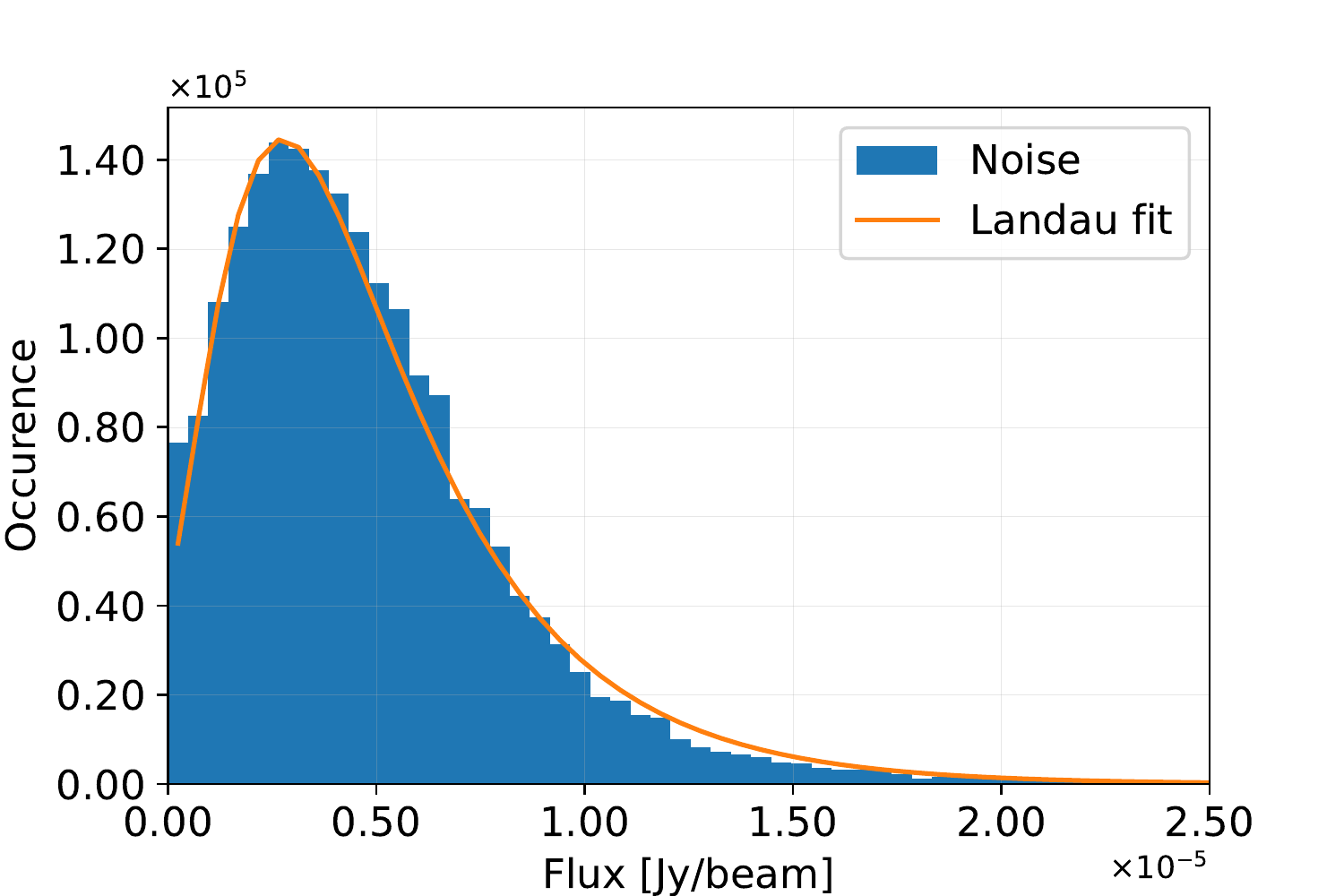}
    \caption{Distribution of noise in one frequency slice of the 2D+1D denoised data cube ($k=3$) fitted with Landau model.}
    \label{fig:landau_noise}
  \end{subfigure}
  \caption{Example of fitting of the noise values in one frequency slice at \SI{995}{\mega\Hz} using two different models for the noisy and 2D+1D denoised data cube. In blue, the histogram of all the voxel value in one frequency slice, i.e. the space plane associated to one frequency. In orange, the fit of these histograms using the two models.}
  \label{fig:noise_fitting}
\end{figure*}

The SDC2 dataset contains Gaussian instrumental noise, RFI noise, and noise approximating the dirty image created by the instrument PSF. We found that the noise in each channel of the development dataset was well-modeled by
the Gaussian probability density function $P_{\rm Gauss}(x, \mu, \sigma)$, where $x$ is the flux measurement. An example fit for the \SI{995}{\mega\Hz} frequency slice is shown in Figure~\ref{fig:gauss_noise}.
While the 2D-1D wavelet denoising removes the vast majority of noise in the dataset, some residual fluctuations still remain. The distribution of these fluctuations takes the form of a skewed Gaussian distribution, and is well-modeled by the approximate Landau distribution:
\begin{equation}
  P_{\rm Landau}(x, s, f) = \frac{1}{f\sqrt{2\pi}}\exp{ -\frac{ (x -s)/f + e^{-(x-s)/f}}{2}}
  \label{eq:landau_approx}
\end{equation}
where $x$ is the flux measurement,  $s$ is a shift parameter, and   $f$ is a scaling parameter.
The result of this fit is shown in Figure~\ref{fig:landau_noise}. These two distributions form the basis of the NHT method.  

Due to frequency-dependent effects such as RFI and variation in the primary beam size, the noise parameters are calculated separately for each frequency channel index $k$, yielding a distinct noise model for each frequency channel with best-fit set of parameters $\alpha_k$, where $\alpha_k$ = ($\mu_k$, $\sigma_k$) in the original data and $\alpha_k$ = ($s_k$, $f_k$) in the denoised data. In a single frequency channel, the likelihood of the null hypothesis $H_0$ given an observation $x_k$ is:

\begin{equation}
  \centering
  \mathcal{L}(H_0 | x_k) = P(x_k |\alpha_k)
  \label{eq:definition_likelihood}
\end{equation}
However, each individual HI source spans more that a single frequency slice.
We use a joint likelihood to construct the null hypothesis for a range of frequency channels:
\begin{equation}
  \mathcal{L}(H_0|x) = \prod_{k=0}^{N} p(x_k |\alpha_k)
  \label{eq:joint_likelihood}
\end{equation}

\begin{figure*}[h!]
  \begin{subfigure}{.5\textwidth}
    \centering
    \includegraphics[width=\textwidth]{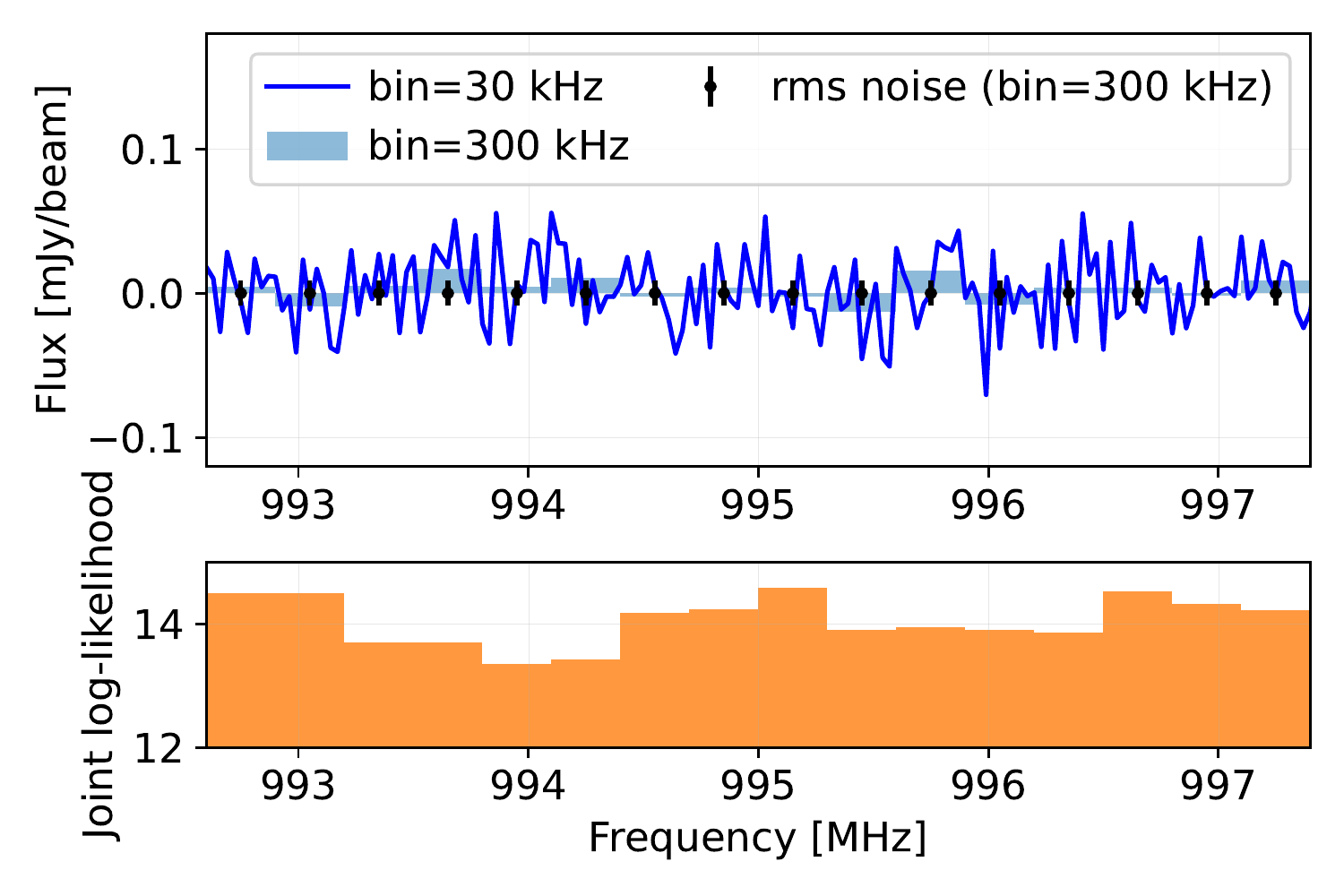}
    \subcaption{Noise sample}
    \label{fig:verif_flux_noise}
  \end{subfigure}
  ~
  \begin{subfigure}{.5\textwidth}
    \centering
    \includegraphics[width=\textwidth]{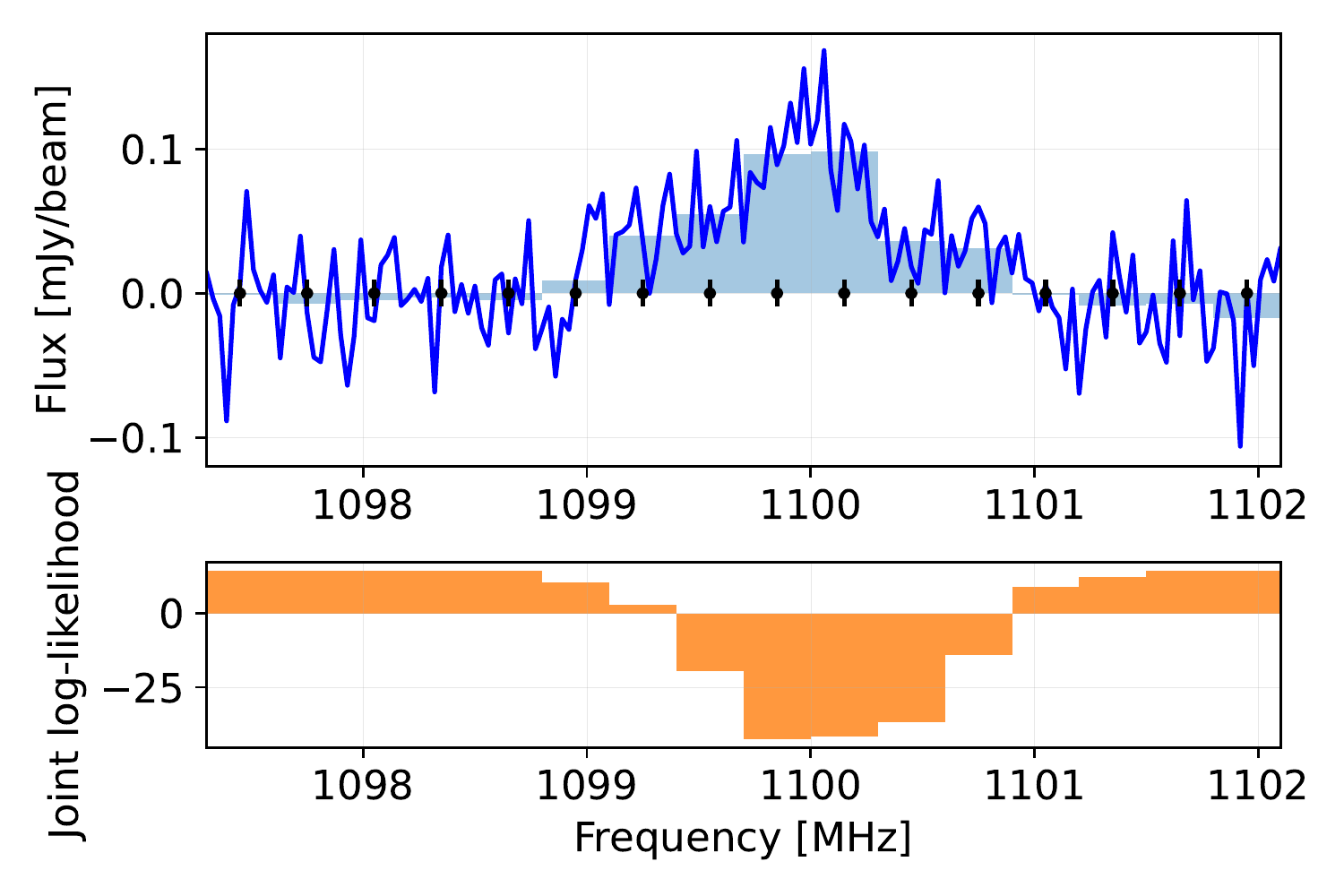}
    \subcaption{Source with integrated line flux above \SI{100}{\Jy\Hz}}
    \label{fig:verif_flux_signal}
  \end{subfigure}
  \caption{Flux and its associated log-likelihood for some frequency range for (\subref{fig:verif_flux_noise}) a random position in space (noise) and (\subref{fig:verif_flux_signal}) a source of integrated line flux above \SI{100}{\Jy\Hz}. In the top plot, one has the flux for binnings of \SI{300}{\kilo\Hz} and \SI{30}{\kilo\Hz}, and the mean and standard deviation of the pdf. At the bottom, the joint likelihood over \SI{1200}{\kilo\Hz} is shown.}
  \label{fig:verif_flux}
\end{figure*}

We choose N so that the joint likelihood is evaluated over $200$ voxels in the SDC2 pipeline, which is much larger than any HI source in the development dataset.
This joint likelihood measures the consistency of the observed data $x_k$ with the expected noise model described by parameters $\alpha_k$ in each channel. If one or more channels have flux values larger than expected from noise, $\mathcal{L}$ will be small. Because $\mathcal{L}$ evaluates the flux relative to expected parameters, it is robust to any channel-dependent effects; a channel with RFI will only have a small $\mathcal{L}$ if the observed flux is much higher than the noise caused by RFI.
%
%
%
$\mathcal{L}$  is used to accept or reject the null hypothesis.
The source finding strategy proceeds as follows:
\begin{enumerate}
\item The frequency channels are combined to improve SNR. We rebin the frequency axis by a factor of 10, which corresponds to the average line width of sources in the development data. This change results in frequency channels with sampling  \SI{300}{\kilo\Hz} or \SI{86}{\km\per\s} at \SI{1050}{\mega\Hz}. 
\item  Noise model parameters $\alpha_k$ are evaluated in each frequency bin $k$.
\item At each point in (RA, Dec), the sliding joint likelihood is calculated along the entire frequency domain (\SIrange{950}{1150}{\mega\Hz}).
\end{enumerate}
The smaller the joint likelihood $\mathcal{L}$, the more likely it is that the data $x_k$ are associated with an HI source instead of noise.
Figures \ref{fig:verif_flux_noise} and \ref{fig:verif_flux_signal} shown the comparison between the flux at a random angular position (i.e. pure noise), the flux for an HI source, and the corresponding likelihoods.
A likelihood threshold $T$ must be chosen such that data $x_k$ with $\mathcal{L} < T$ are HI sources. Thus the final step of source finding is:
\begin{enumerate}[resume]
  \item If one of the joint likelihoods is smaller than some threshold $T$, the null hypothesis will be rejected and the position will be considered as a source candidate.
\end{enumerate}

The choice of this threshold drastically impacts the performance of the source finding procedure.
\texttt{LiSA} contains a function to evaluate this $T$ automatically using the development dataset. It evaluates all of the likelihoods $\mathcal{L}$ on random sections of the data cube, which are highly likely to be noise, plots the distribution of likelihoods, and chooses a threshold which should eliminate 99.7\% of false sources. However, users may also supply their own choice for $T$ depending on their desired performance.

\subsection{True and False Positive Rates}
For the full SDC2 pipeline, $T$ is chosen by evaluating the Receiver Operating Characteristic (ROC)
of the source finder on the development dataset. We evaluated the performance of the source finding procedure on both the original and denoised data. All NHT parameters are kept the same except for the  function choice.

\begin{figure*}[h!]
  \begin{subfigure}{.5\textwidth}
    \centering
    \includegraphics[width=\textwidth]{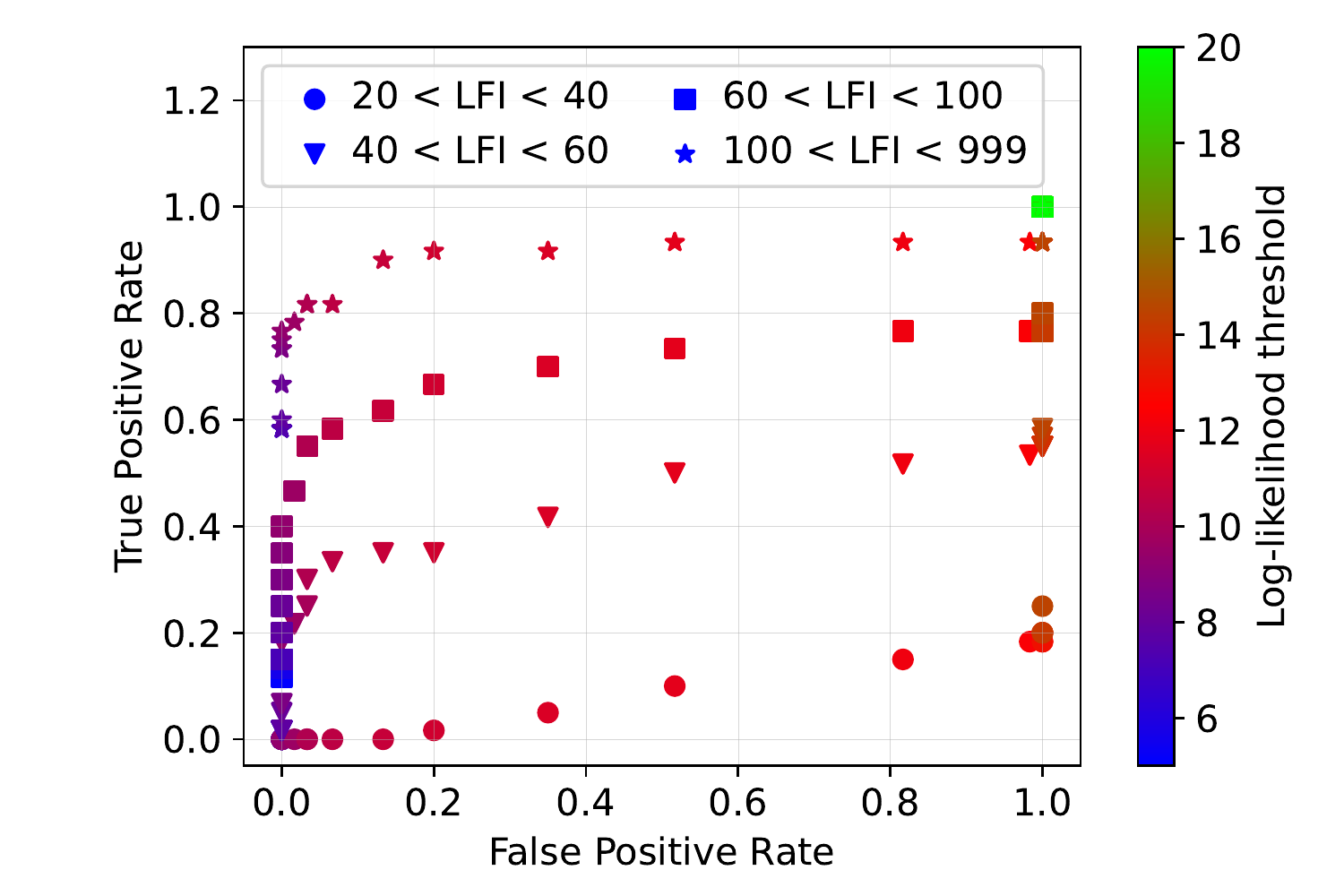}
    \caption{Noisy data cube}
    \label{fig:roc_bin10_binary}
  \end{subfigure}
  ~
  \begin{subfigure}{.5\textwidth}
    \centering
    \includegraphics[width=\textwidth]{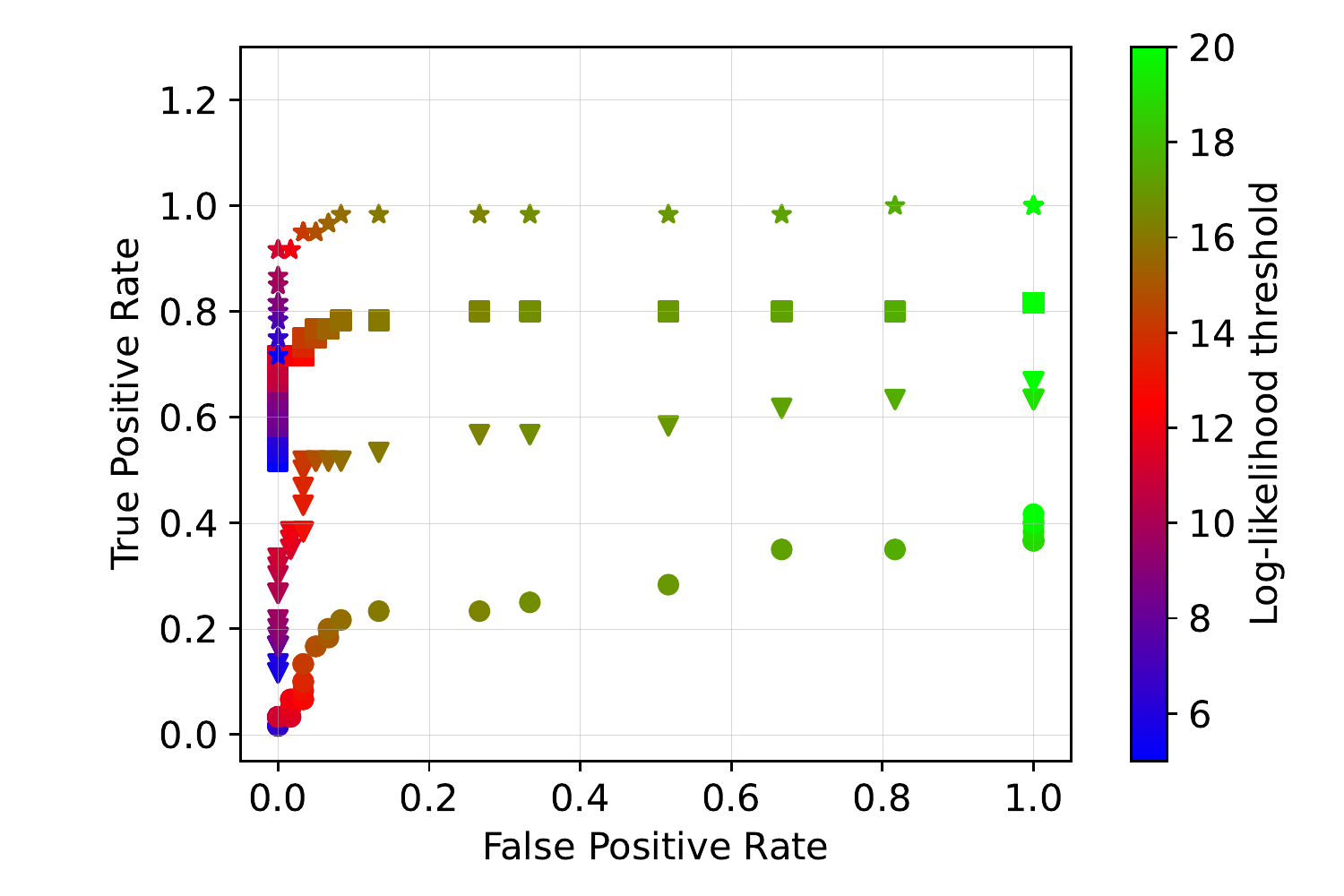}
    \caption{Wavelet-denoised data cube}
    \label{fig:roc_bin10_astro}
  \end{subfigure}
  \caption{ROC curve for the denoised and original data. The frequency axis in both data cubes is rebinned by  a factor of 10.
  For different ranges of the HI source integrated line flux (LFI), the true positive and false positive rate of the null hypothesis test is evaluated for a range of thresholds $T$ uniformly distributed between \SIrange{5}{20}{}.
 LFI values are in units of Jansky Hertz.
}
  \label{fig:roc_bin10}
\end{figure*}

Figure \ref{fig:roc_bin10} shows the NHT performance for both the datasets for a variety of threshold choices and different HI source brightnesses.
The performance is strongly improved by denoising the dataset.
The true positive rate for the sources in the brightness ranges above \SI{60}{\Jy\Hz} were both improved to a perfect classification.
For dimmer sources with integrated line flux from \SIrange{20}{60}{\Jy\Hz}, the results also increased, the true positive rate improved from less than \SI{10}{\percent} to about \SI{40}{\percent} while keeping the same \SI{10}{\percent} false positive rate.
Finally for the sources with brightness under \SI{20}{\Jy\Hz}, no improvement is observed.
In the end, the denoising process shows an overall improvement of \SIrange{10}{20}{\percent} for the true positive rate at equivalent false positive rate. 
The final pipeline used a threshold of $T=12$ on the denoised dataset as, based on the ROC curves, it corresponds to the elbow of the curves .i.e. high true positive rate for low false positive rate.
Different threshold were tested. 
Decreasing it improves the purity of the catalogue, at the cost of less sources found.
On the other hand, increasing it will reduce the purity while increasing the number of sources in the catalogue.
The choice of $T=12$ for the threshold is reasonable compromise.

\subsection{Comparison to SoFiA}

We also compare the performance of the null-hypothesis source finder to the 
 Source Finding Application SoFiA (\cite{SoFiA}), which was used by several other teams in the SDC2 challenge. We ran the SDC2 SoFiA team's pipeline\footnote{https://github.com/SoFiA-Admin/SKA-SDC2-SoFiA}, but without the additional cleaning step to remove false positives to have a direct comparison of the source finding algorithms.
 
\begin{figure}[h!]
    \centering
    \includegraphics[width=0.45\textwidth]{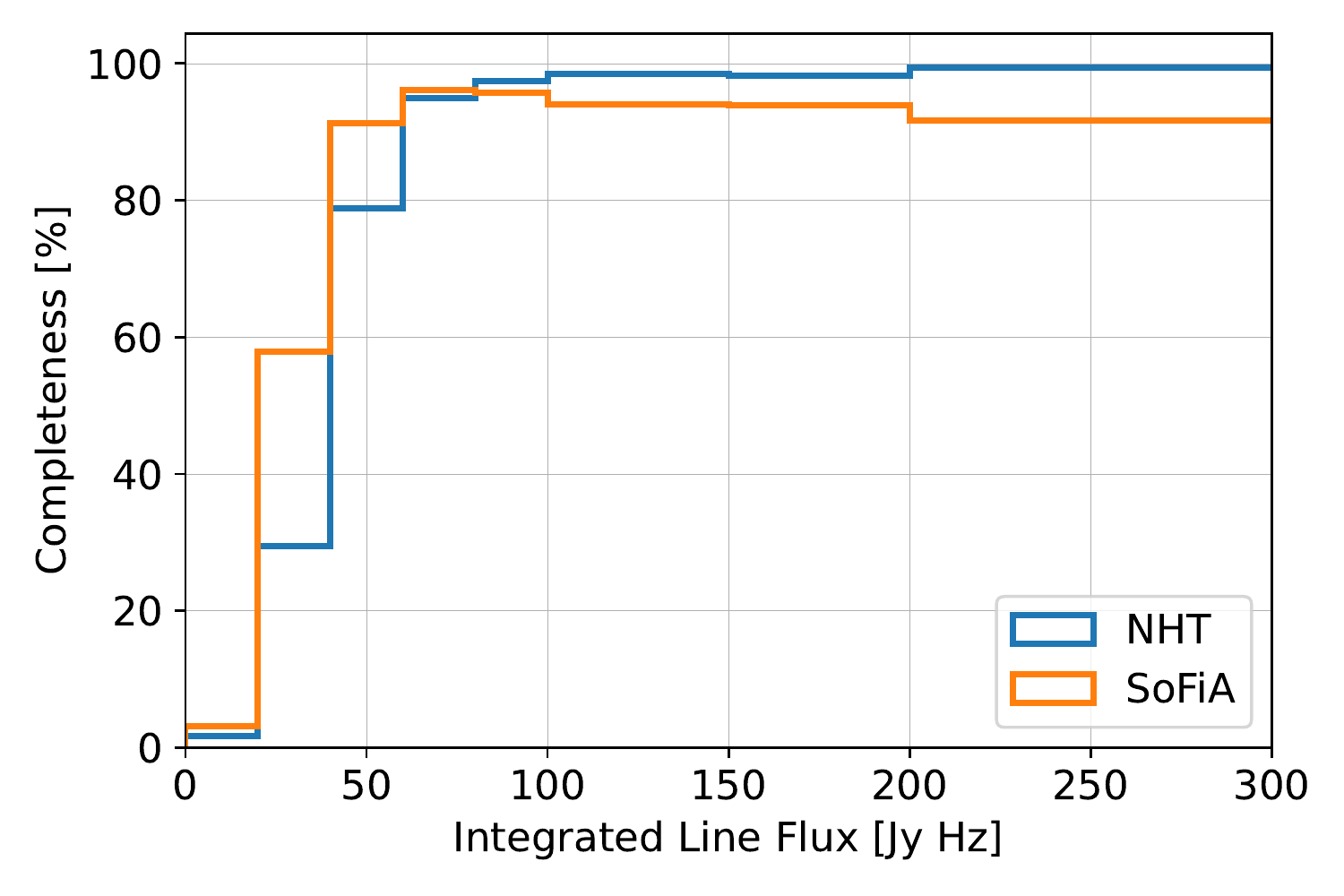}
    \label{fig:compl_bin10_raw}

  \caption{Comparison of the completeness for SoFiA on the original data and the null hypothesis test on the denoised data cube. The NHT results are computed for a threshold of 12.}
  \label{fig:compl_bin10}
\end{figure}

Figure \ref{fig:compl_bin10} shows the comparison of completeness for different brightness ranges.
The null hypothesis finding has better completeness at higher brightness, but the performance decreases with respect to SoFiA at lower brightness.
Regarding the total completeness, the NHT is at \SI{9.14}{\percent} and SoFiA is at \SI{13.93}{\percent}, due to much more sources at lower brightness as shown in Figure~\ref{fig:var}.

\section{Source Filtering \& Characterization}

\subsection{Convolutional Neural Networks}

Deep neural networks are well known for their capability as universal function approximators, and have often been used to accelerate or automate data analysis. Convolutional neural networks  (CNN, \cite{cnn}) have been increasingly used in the field of astronomy over the past decade for the classification of 2D data, to recognize galaxy morphologies \citep{astronn1}, identify galaxy-scale strong gravitational lenses \citep{astronn2}, or find radio sources in the sky \citep{astronn3}.
\texttt{LiSA} contains two CNNs that operate on 3D input data and were trained on the SDC2 development dataset:
\begin{enumerate}
\item A {\bf classification} network used to filter true and false HI sources identified by the NHT step
\item A {\bf regression} network used to predict the HI source properties
\end{enumerate}

The SDC2 development dataset and associated truth catalog were used to generate training and validation data for both networks. However, as the truth catalog only contains approximately 10,000 sources, extensive data augmentation (\cite{ShortenK19}) and regularization were needed to prevent overfitting during network training. We extensively use  Dropout  (\cite{dropout})  to regularize the learning of the network.

\subsection{Source Filtering}

Source candidates are generated by applying the NHT procedure to the entire  data domain, assigning a likelihood score to every voxel. Voxels below the user-defined threshold are discarded. Of the remaining voxels $V$, neighbors are merged using an iterative island-finding algorithm: 1) every above-threshold voxel is assigned to an island, 2) if any two islands have voxels that are closer than 10 pixels, merge the islands, and 3) repeat until none of the islands change. Each island returns the location of the most-significant voxel as the source candidate position.

\begin{figure}[h!]
    \centering
    \includegraphics[width=0.5\textwidth]{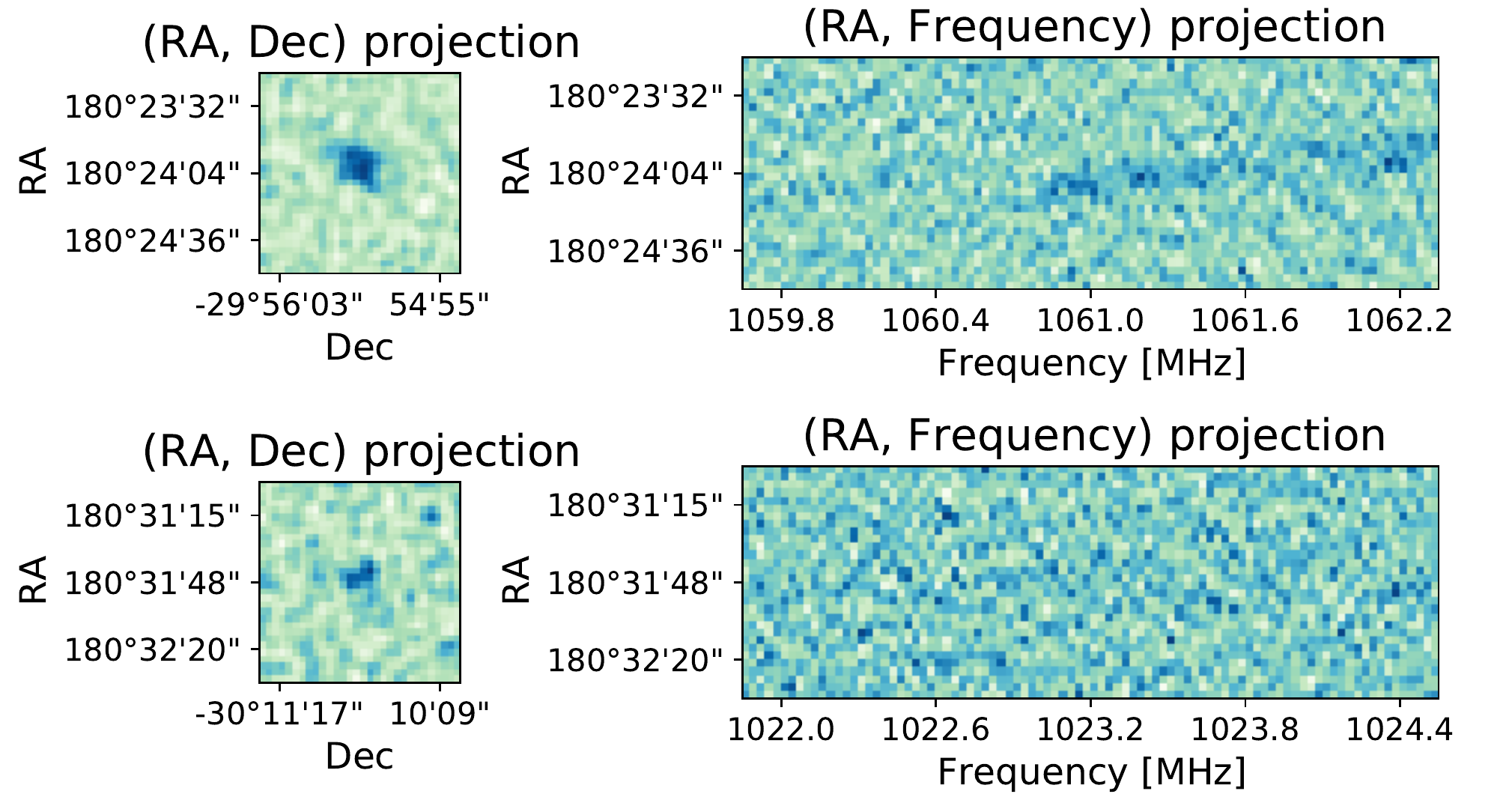}
  \caption{Cutouts in (Ra, Dec, Freq) around a true HI source (top row) and false HI source (bottom row). The cubes are projected along the z (Frequency) axis (left column) and along the x (RA) axis (right column). \label{fig:fakedata}}
\end{figure}

However, the dataset also contains several data artifacts arising from imperfect continuum subtraction which also pass the null hypothesis test, shown in Figure~\ref{fig:fakedata}. To separate these artifacts from true HI sources we developed a classifier CNN.

To generate labelled training data, we ran the denoising and source finding steps of the pipeline on the development dataset.
We then cross-matched the resulting source ~candidates ~with ~the ~development
dataset truth catalog to create a sample of 1249 true HI sources and and 3589 false HI sources. Subcubes  with dimensions (30,30,200) in (RA, Dec, Freq) centered around each true and false source were extracted from the development dataset to generate training and validation data. Training data was augmented by rotations of $90^{\circ}$, $180^{\circ}$, or $270^{\circ}$ in the (RA, Dec) plane, reflections in the (RA, Dec) plane, small random translations of 0 to 5 voxels along each axis, and scaling the flux by a coefficient randomly sampled between $0.9$ and $1.3$. No augmentations were applied to validation data.

\begin{figure*}[h!]
    \centering
    \includegraphics[width=0.99\textwidth]{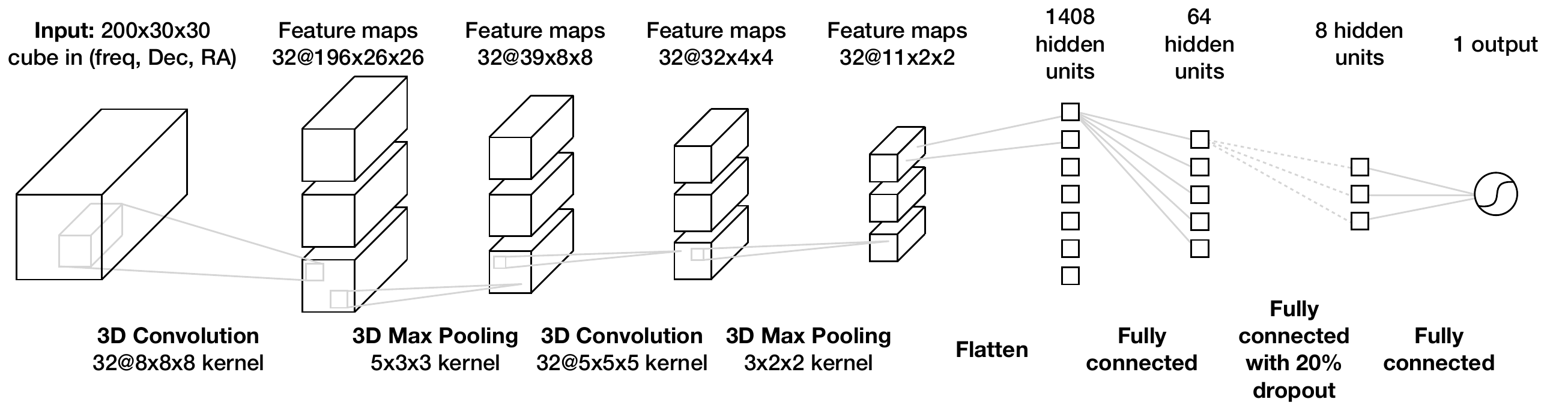}
  \caption{Architecture of the classifier convolutional neural network. All nodes use rectified linear activation functions except for the final output layer which uses a sigmoid activation function.\label{fig:classarc}}
\end{figure*}

The network architecture is shown in Figure~\ref{fig:classarc}.
This network was trained for 10 epochs with a batch size of 64 using the RMSProp optimizer with
binary crossentropy loss.

\begin{figure}[h!]
    \centering
    \includegraphics[width=0.4\textwidth]{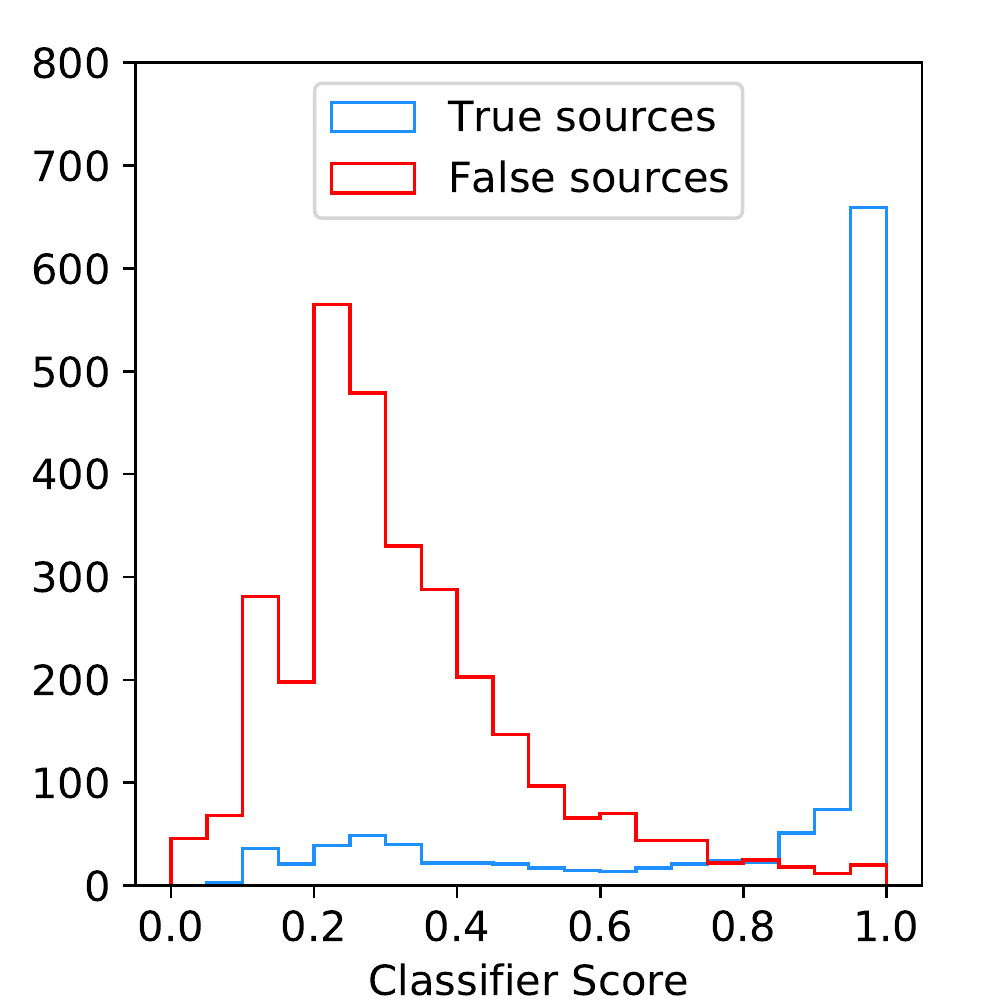}
  \caption{The classifier score (Right) assigned to fake validation data (red) and true validation data (blue). \label{fig:classifierperf}}
\end{figure}

The final trained network is able to separate the true and false HI sources with high accuracy, shown in Figure~\ref{fig:classifierperf}. In the full pipeline, subcubes around each source candidate are passed to the network and assigned a score. All source candidates with a score below  $0.85$ are discarded, which keeps 94\% of true sources and discards 89\% of false sources. 
The remaining sources form the basis of the source catalog and are passed to the source characterization step of the pipeline. 

\subsection{Source Characterization}

Once the final source candidate locations have been selected, several parameters of the HI sources such as line flux integral and HI size are calculated. We defined a neural network that takes in a subcube around a source candidate location with dimensions (30,30,200) in (RA,Dec,Freq) and outputs the five parameters shown in Figure~\ref{fig:var}. To normalize these data distributions we rescaled each of the HI source parameters to lie between 0 and 1. The neural network is trained to predict these transformed parameters. Furthermore, position angle is a cyclical variable and data look similar for $\theta \sim 0$ and $\theta \sim 360$. Instead of setting up the network to predict $\theta$, we instead train the network to predict $\sin \theta$ and  $\cos \theta$.

\begin{figure}[h!]
  \centering
    \includegraphics[width=0.4\textwidth]{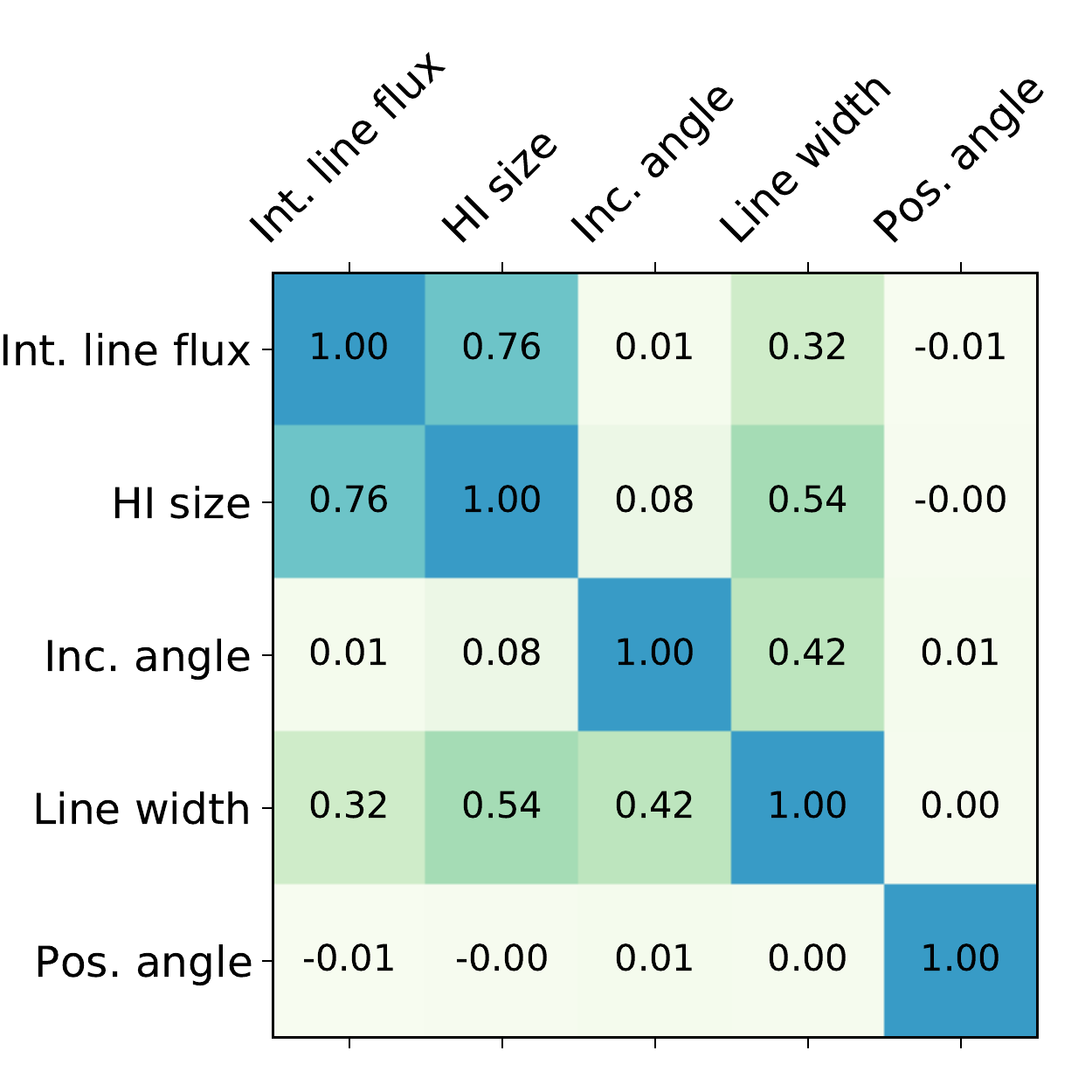}
  \caption{Correlation matrix for the different HI source properties in the development dataset. Several of the parameters have covariance, especially line flux integral and HI size. However position angle exhibits no correlation with any other variable.\label{fig:covar}}
\end{figure}

The exponential distribution of many of the variables makes this source characterization an imbalanced regression problem. There are comparatively few HI sources that are bright, large, or with small inclination angle.
To counteract this effect we implemented a custom data generator that begins by sorting all instances of the training data into different equally-spaced 20 Jy Hz ranges based on their true flux $F$: $0 < F < 20$ Jy Hz, $20 < F < 40$ Jy Hz, up to $F > 200$ Jy Hz. To generate the batch of data for training, the generator randomly chooses sources from each of these bins with equal probability. 
Because line flux integral is highly correlated with HI size and line width, increasing the probability to see brighter sources also increases the probability to see larger sources. However, as there are fewer than 20 sources in the highest generator bin, the network is susceptible to overfitting.

To prevent overfitting, the training data are subjected to extensive data augmentation including random rotations between  $90^{\circ}$ and $270^{\circ}$ in the (RA, Dec) plane,
reflections in the (RA, Dec) plane, small random translations of $0-5$ voxels along each axis, and scaling the flux by a coefficient randomly sampled between $0.9$ and $1.3$. Because all training data are rotated by a random angle between $90^{\circ}$ and $270^{\circ}$, we use unaugmented and unrotated data as our validation sample. 
In addition to data augmentation, the network also includes multiple dropout layers. Dropout randomly removes network connections during training, disrupting which features can be seen by deeper layers of the network. 

The data associated with bright galaxies has a better signal to noise, allowing the network can learn features more easily from these galaxies. Additionally, the previous source-finding steps of the pipeline are more likely to discover bright galaxies, so it is especially import for the network to predict their properties correctly. To increase penalties for incorrect characterization of bright galaxies, and therefore improve the performance of this characterization, we implemented a custom loss function which rescales the mean square error loss proportionally to the flux of the true source:

\begin{equation*}
    L(y, \hat y) = \frac{1+\hat y_0}{N} \sum_{i=0}^N (y_i-\hat y_i)^2
\end{equation*}
where $y$ is the vector of $N$ true source properties, $\hat y$ is the vector of $N$ predicted source properties, and $\hat y_0$ is the predicted line flux integral. 

\begin{figure*}[ht!]
    \centering
    \includegraphics[width=0.99\textwidth]{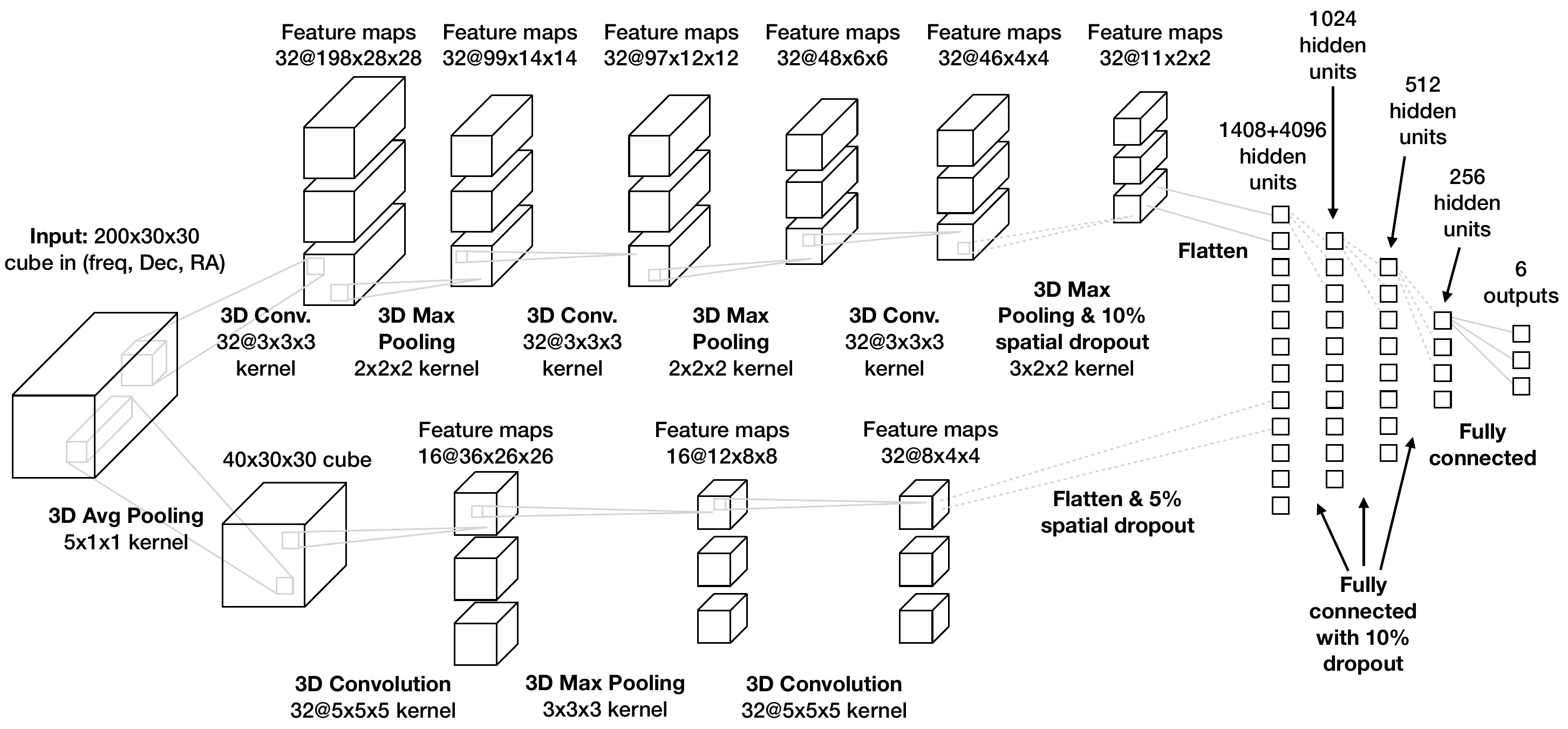}
  \caption{Architecture of the regression convolutional neural network. All nodes use rectified linear activation functions except for the final output layer which uses linear activation functions. \label{fig:classreg}}
\end{figure*}

The network is designed to analyse features in 3D spectral data. However, the scales of features in spatial and spectral dimensions differ greatly. As such, we decided to implement two different parallel convolutional blocks which use different-sized filters to analyse features of the data cube at different scales. These features are then concatenated into one vector, which is then processed by fully connected layers to calculate the source parameters.
This multi-branched structure enhances the granularity and robustness of the network, and was inspired by the inception model introduced in~\cite{incep} and used for galaxy morphology classification in \\ \cite{galincep}.
The regression network architecture is shown in figure~\ref{fig:classreg}

\begin{figure*}[h!]
    \centering
    \includegraphics[width=0.99\textwidth]{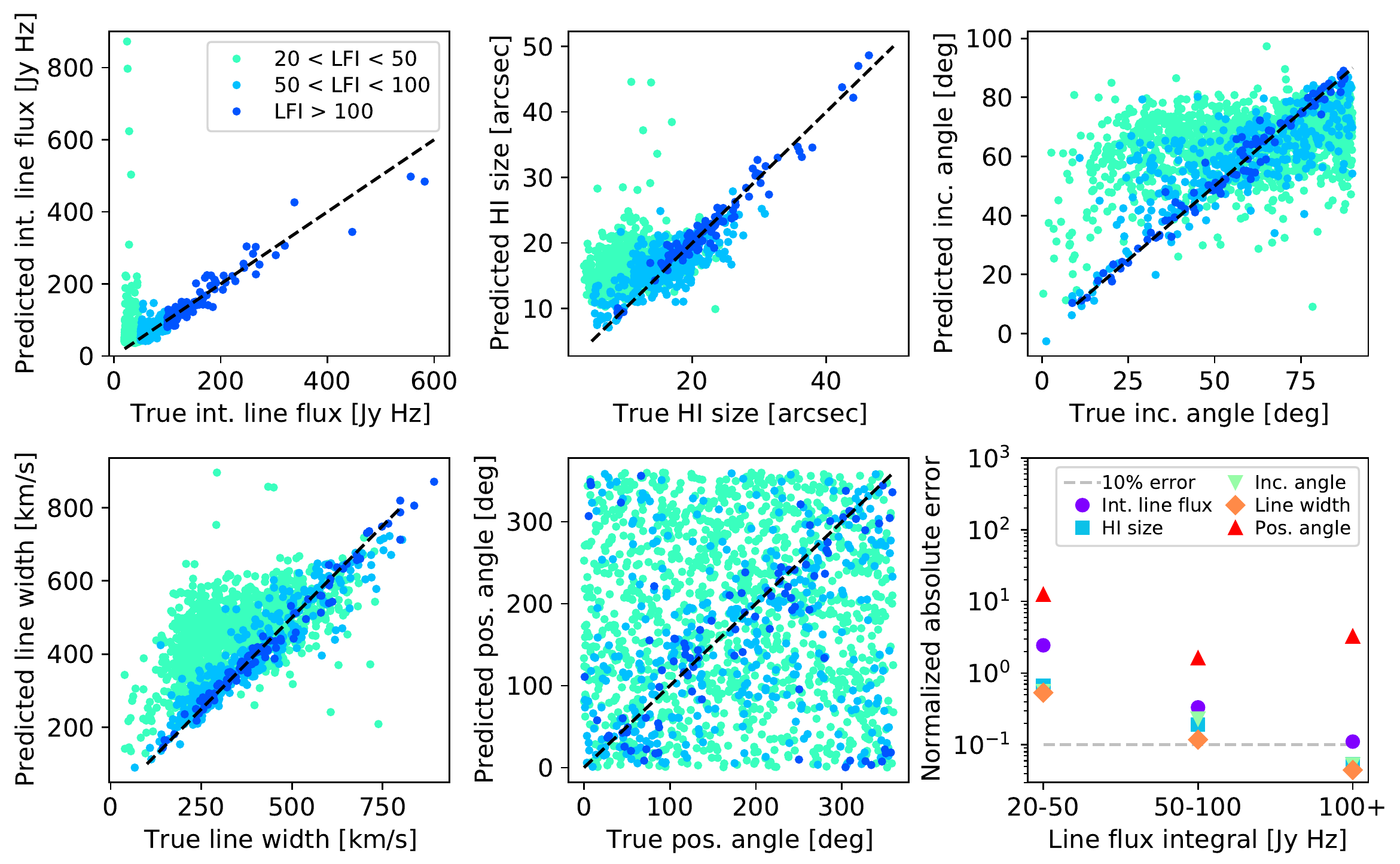}
  \caption{Performance of the regression CNN. The network can reliably predict the HI source properties to within 10\% for the brighter sources in the catalog with integrated line flux $ > 50 $, which corresponds to SNR $\gtrsim 3 $. The network cannot reliably predict the properties of dimmer sources with SNR $ < 3$.  \label{fig:regperf}}
\end{figure*}

The regression network was trained for 300 epochs with a batch size of 32 using the Adam optimizer with a learning rate of $10^{-4}$. Network predictions for the the five variables are shown in Figure~\ref{fig:regperf}. We observe good accuracy for line flux integral, HI size, inclination angle, and line width. As The network struggles to make meaningful predictions for the position angle variable for all but the very brightest sources. This is due to the correlation between the different parameters as shown in Figure~\ref{fig:covar}. All of the parameters exhibit some covariance except for position angle. This means that while observables like the brightness and extent of the galaxy can inform the network's prediction for line width, the network has fewer constraints on the position angle prediction.

In the final pipeline, subcubes from the original data are extracted at each of the source candidate locations. These subcubes are passed to the network, which returns values for the five source parameters. These five parameters and the location information for each source form the HI source catalog entry.
The HI source locations and features for each domain are then concatenated to create the full catalog.

\section{Summary}
We have shown the performance of the different \texttt{LiSA} modules on the SDC2 dataset. 
The source finding modules are designed to be flexible and generalizable with as few user-defined inputs as possible:
\begin{itemize}
    \item The \textbf{domain decomposition} step only takes in the number of domains and the border size as input arguments. The number of domains must be chosen based on the system memory requirements, and the border size should be selected depending on the maximum size of expected objects in the dataset.
    \item The \textbf{wavelet denoising} only needs the regularization strength $\lambda$, which can be defined in terms of signal-to-noise such as $3\sigma$ detections, $5\sigma$ detections, etc.
    \item The \textbf{null-hypothesis testing} automatically evaluates the noise parameters of the dataset. The user must simply choose a function to represent the noise, a likelihood threshold value, and two parameters to represent the channel resolution. The code can choose the threshold value using properties of the noise, or the user can supply a value manually.
\end{itemize}

The CNN source filtering and source characterization steps are not as portable. The networks were trained on the specifics of the SDC2 data, and cannot be used out-of-the box on different datasets.
However, the trained networks are also publicly available as part of \texttt{LiSA}, and with transfer learning \citep{transferlearning} they can be adapted for other 3D datasets with a small amount of training data. This procedure was applied sucessfully for the 2D galaxy morphology classification network \textsc{claran}  \citep{claran}, which was trained on on the \textsc{FIRST} and \textsc{WISE} images from the Radio Galaxy Zoo Data Release 1 catalogue\footnote{https://data.galaxyzoo.org/} and adapted for the SKA Science Data Challenge 1 dataset\footnote{https://github.com/ICRAR/skasdc1} \citep{sdc1}.

\begin{figure*}[h!]
  \begin{subfigure}{.5\textwidth}
    \centering
    \includegraphics[width=\textwidth]{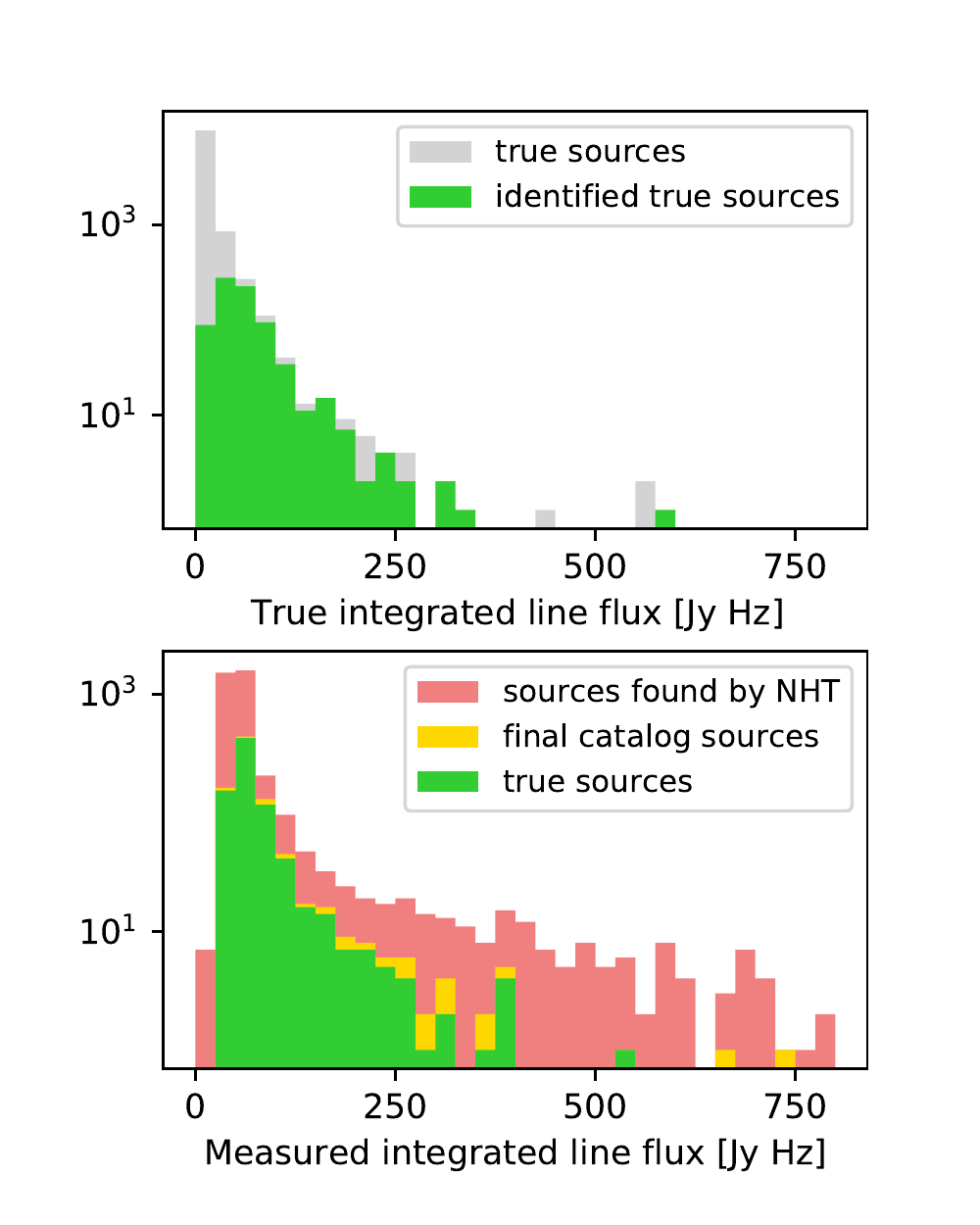}
    \subcaption{Performance on the SDC2 development datacube}
    \label{fig:dev_perf}
  \end{subfigure}
  ~
  \begin{subfigure}{.5\textwidth}
    \centering
    \includegraphics[width=\textwidth]{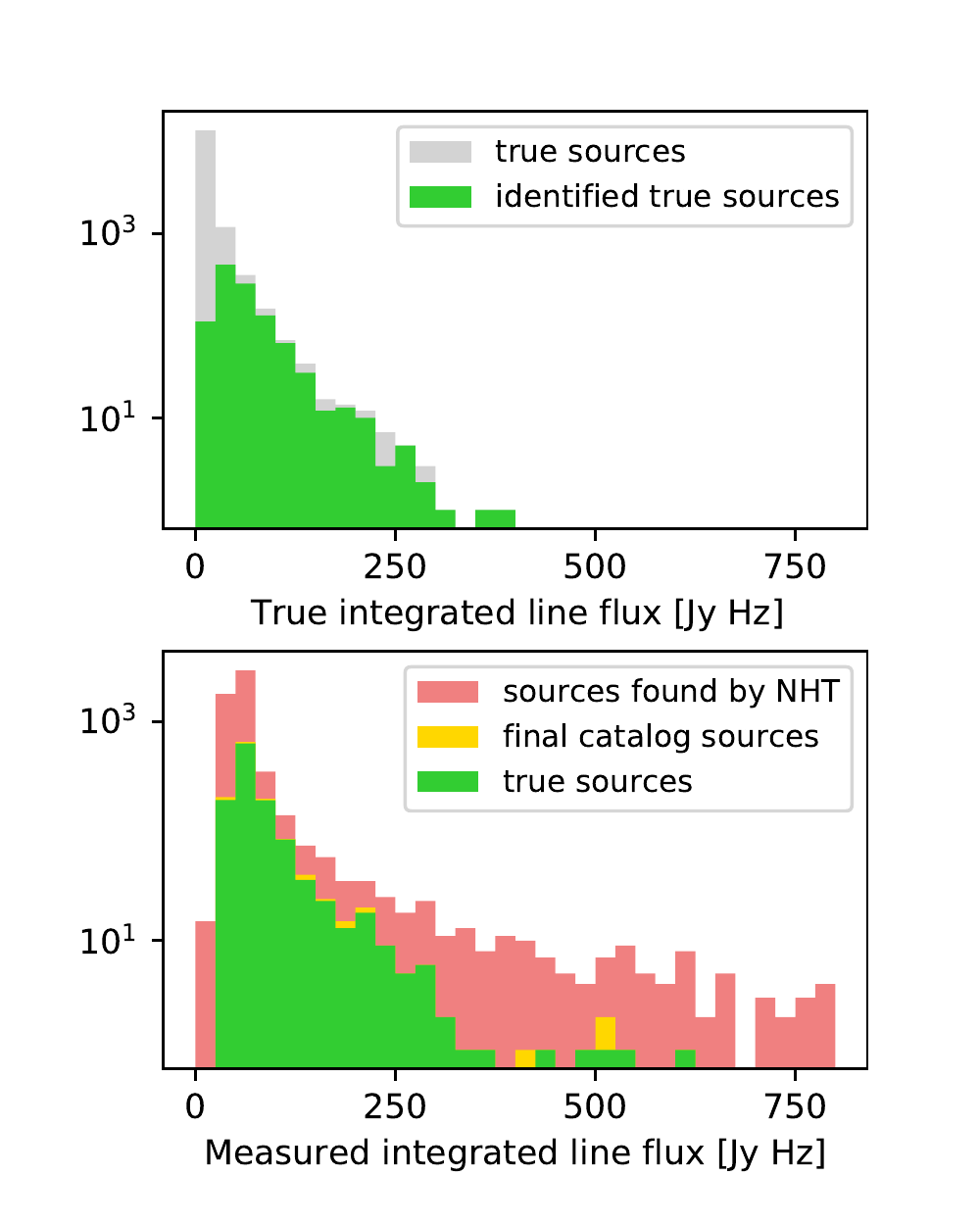}
    \subcaption{Performance on the SDC2 full datacube}
    \label{fig:full_perf}
  \end{subfigure}
  \caption{Performance of the \texttt{LiSA} pipeline on the SDC2 development dataset (left) and full dataset (right). The deep learning components of the pipeline were trained on data-augmented cutouts of the development dataset, but were completely blind to the full dataset. We observe consistent performance between both datasets. In the above plots, a source in the catalog is considered matched to a true source if they are within 30 arcsec and 50 kHz.}
  \label{fig:pipeline_perf}
\end{figure*}

The library also contains two pipeline for running on the SDC2 development and full datasets, which demonstrate the domain decomposition and parallel processing, and all of the algorithms described in this paper. The performance of the pipeline on these data are excellent as shown in Figure~\ref{fig:pipeline_perf}, and can consistently identify sources with integrated line flux $> 50$ Jy Hz.

The modularity and minimal tunable parameters of this software will allow users to
optimize their source-finding and characterization strategy on their datasets of interest.

\section{Acknowledgements}
MTS acknowledges support from a Scientific Exchanges visitor fellowship (IZSEZO\_202357) from the Swiss National Science Foundation. 
\appendix

\bibliographystyle{elsarticle-harv} 
\bibliography{cas-refs}

\end{document}